\documentstyle[12pt,epsf,aasms4]{article}

\def\ni{\noindent}
\def\pj{\hspace{-.26cm}}
\def\fpj{\hspace{-.7cm}}

\def\thalf{{\textstyle{\frac{1}{2}}}}

\begin{document}

\title{ NEUTRINO SCATTERING IN A NEWLY BORN NEUTRON STAR }  
\author{ SANJAY REDDY AND MADAPPA PRAKASH } 
\affil{Physics Department \\ 
      	State University of New York at Stony Brook\\ 
       	Stony Brook, NY 11794-3800, USA }

\begin{abstract}

We calculate neutrino cross sections from neutral current reactions  in the 
dense matter encountered in the evolution of a newly born neutron star. 
Effects of composition and of strong interactions in the deleptonization and
cooling phases of the evolution are studied. The influence of the possible
presence of strangeness-rich hyperons on the neutrino scattering cross sections
is explored.   Due to the large vector couplings of the $\Sigma^-$ and $\Xi^-$,
$|C_V| \sim 2$, these particles, if present in protoneutron star matter, give
significant contributions to neutrino scattering.  In the deleptonization
phase, the presence of strangeness leads to large neutrino energies, which
results in large enhancements in the cross sections compared to those in matter
with nucleons only.  In the cooling phase, in which matter is nearly neutrino
free, the response of the $\Sigma^-$ hyperons to thermal neutrinos is the most
significant.   Neutrinos couple relatively weakly to the $\Lambda$ hyperons 
and, hence, their contributions are significant only at high density. \\

\end{abstract}            

\section{ INTRODUCTION}

The general nature of the neutrino signature expected from a newly formed
neutron star (hereafter referred to as a protoneutron star) has been
theoretically predicted~(Burrows \& Lattimer 1986) and  confirmed by the 
observations~(Bionta et al. 1987; Hirata et al 1987)  from supernova SN1987A. 
Although neutrinos interact weakly with matter, the high baryon densities and
neutrino energies achieved after the gravitational collapse of a massive star
($\geq 8$ solar masses) cause the neutrinos to become trapped on the dynamical
time scales of collapse~(Sato 1975; Mazurek 1975).  Trapped neutrinos at the
star's core have  Fermi energies $E_\nu \sim 200-300$ MeV and are primarily of
the $\nu_e$ type. Thermally produced $\mu$ and $\tau$ neutrino pairs are also
trapped, but with zero chemical potential.  Neutrinos escape after diffusing
through  the star exchanging energy with the ambient matter, which has an
entropy per  baryon of order unity in units of Boltzmann's constant. 
Eventually they  emerge from the star with an average energy $\sim 10-20$ MeV
and in nearly  equal abundances of all three flavors, both particle and
anti-particle.

Neutrino interactions in dense matter have been investigated by various
authors (Tubbs \& Schramm 1975; Sawyer 1975,89,95; Lamb \& Pethick 1976; 
Lamb 1978; Sawyer \& Soni 1979; Iwamoto \& Pethick 1982; Iwamoto 1982; 
Goodwin \&Pethick 1982; Burrows \& Mazurek 1982; Bruenn 1985; 
van den Horn \& Cooperstein 1986; Cooperstein 1988; Burrows 1988; Horowitz
and Wehrberger 1991a,b;1992; Reddy \& Prakash 1995). 
The charged current absorption and neutral current scattering reactions are 
both important sources of opacity. The neutral current scattering involves all
flavors of neutrinos scattering on nucleons and leptons. Scattering from 
electrons is important for energy and momentum transfer (Tubbs and Schramm, 
1975).  However, for lepton number transport, 
nucleon scattering and absorption are the dominant processes. 

Surprisingly little attention has been paid to the effects of composition and
of strong interactions of the ambient matter on neutrino opacities.  In the few
attempts to date, the effect of interactions was investigated for
non-degenerate nuclear matter by Sawyer  (1975; 1989) and for degenerate pure
neutron matter by Iwamoto \& Pethick  (1982). Treating nucleons in the
non-relativistic limit, these calculations  predict an increase in the mean
free path by a factor of  $\sim 2-3$, for  (2-4) times the nuclear density. 
More recently, relativistic calculations  based  on  effective Lagrangian
models for dense neutron star  matter have  been performed by Horowitz and
Wehrberger (1991a,b;1992).  Here, the  differential cross-sections for matter 
containing nucleons and  electrons  were  calculated using linear response
theory.  A reduction of 30-50\% over  the case of non-interacting nucleons was
reported in these calculations.    In addition to strong interaction
modifications,  electromagnetic interactions can increase the mean free path by 
50-60\% for electron type neutrinos through collective effects 
(Horowitz 1992).   So far, the influence of interactions has been investigated
in protoneutron star  calculations only by a simple scaling of the
non-interacting results  (Burrows 1990, Keil 1994).  Furthermore, there have
been no calculations  performed including the multi-component nature of the
system.  We note that  Keil and Janka (1995) have recently carried out 
deleptonization  and cooling
simulations including hyperons in the EOS, but they
ignored opacity  modifications.  We view it as essential that opacities be
consistent with the composition, which has not been a feature of protoneutron
star models to date.  The composition of the star changes significantly from
the deleptonization phase,  in which neutrinos are trapped, to the cooling
phase, in which only thermally produced neutrinos are present.  So far,
important opacity modifications due to the changing lepton content and
composition have not been treated satisfactorily in evolutionary 
calculations.              
         
Although the composition and the equation of state (EOS)  of the hot
protoneutron star matter are not yet known with certainty, QCD based effective
Lagrangians have opened up intriguing possibilities (Kaplan \& Nelson (1986);
Glendenning 1986,1992;  Glendenning \& Moszkowski (1991); Kapusta \& Olive
(1990); Ellis, Knorren \&  Prakash (1995); Knorren, Prakash \& Ellis (1995),
Prakash, Cooke \& Lattimer 1995). Among these is the possible existence  of
matter with strangeness to baryon ratio of order unity.  Strangeness may be
present either in the form of fermions, notably the $\Lambda$ and $\Sigma^-$
hyperons, or, in the form of a Bose condensate, such as a $K^-$- meson
condensate, or, in the form of $s$ quarks.   In the absence  of trapped
neutrinos,  strange particles are expected to appear around $2-4$  times the
nuclear matter density of $n_0=0.16~{\rm fm}^{-3}$.   Neutrino-trapping causes
the  strange particles to appear at somewhat  higher densities, since the
relevant chemical potential $\mu = \mu_e -\mu_{\nu_e}$ in matter with high
lepton content is much smaller than in the untrapped case (Thorsson, Prakash \&
Lattimer 1994; Ellis, Knorren \& 
Prakash 1995; Knorren, Prakash \& Ellis 1995, Prakash et al 1996).   
                        
A new feature that we consider here is the role of strangeness.  To date, only
neutrino opacities for strange quark matter have been calculated (Iwamoto,
1982).  Here, we study neutrino scattering mean free paths in matter containing
strangeness in the form of hyperons.  Specifically, we calculate neutrino
opacities from neutral current reactions  in matter containing hyperons  and
which are faithful to the EOS. In a first effort, this will be achieved using a
mean field theoretical description which includes hyperonic degrees of freedom. 
This approach has several merits.  For example,  aspects of relativity, which
may become important at high density, are  naturally incorporated. 
Modifications of the opacity due to correlations (RPA) are also possible in 
such an approach.   Further, comparisons with alternative potential model 
approaches (Iwamoto \& Pethick, 1982; Sawyer, 1989) are straightforward.
Neutrino opacities in matter containing other forms of strangeness  will be
considered in a separate work.  Contributions from charged current reactions
are essential for a complete description of the protoneutron star
evolution.   With appropriate modifications of the formalism presented in this
work, calculations that include the compositional changes in the distinct
phases of the evolution will be reported in a later work.  
       
In \S 2, the formalism to calculate the neutrino scattering cross sections  in
a multi-component system is discussed.  New analytical formulae for the
response functions, which
facilitate  accurate calculations of the cross sections in 
all regimes of matter degeneracy are derived. In \S 3, the composition of
beta-equilibrated matter with and without strange baryons is determined based
on a field theoretical model.  This section also contains a description of the 
relevant physical conditions in the evolution of a newly born neutron star. 
In particular, the composition in the distinct phases of deleptonization and
cooling are discussed.  Results of the neutrino scattering cross sections in
these two  phases are presented in \S 4.  Conclusions and directions for future
study are given in \S 5.

\section{  NEUTRINO INTERACTIONS WITH BARYONS }

Neutrino interactions with matter proceed via charged and neutral
current reactions. The neutral current processes contribute to
elastic scattering, and charged current reactions result in neutrino
absorption.  The formalism to calculate neutral current scattering rates in
dense matter is summarized below.  
The interaction Lagrangian for neutrino
scattering reactions is based on the Wienberg-Salam-Glashow
 theory (Weinberg 1967;
Salam 1968; Glashow 1961):
\begin{eqnarray}
{\cal L}_{int}^{nc} &=& ({G_F}/{2\sqrt{2}}) ~~l_\mu
j_z^\mu\, \qquad {\rm ~for} \qquad \nu + B \rightarrow  \nu + B \,,
\end{eqnarray}
where $G_F\simeq 1.436\times 10^{-49}~{\rm erg~cm}^{-3}$ is the weak coupling
constant.  The neutrino and target particle weak neutral currents appearing 
above are:
\begin{eqnarray}
l_\mu^\nu &=& {\overline \psi}_\nu \gamma_\mu
\left( 1 - \gamma_5 \right) \psi_\nu \,, \nonumber\\
j_z^\mu &=& {\overline \psi}_i \gamma^\mu
\left( C_{Vi} - C_{Ai} \gamma_5 \right) \psi_i \,,
\end{eqnarray}
where $i=n,p,\Lambda,\Sigma^-,\Sigma^+,\Sigma^0,\Xi^-,\cdots$ and 
$e^-,\mu^-$.  The
neutral current process couples neutrinos of all types ($e,\mu$, and $\tau$) to
the  weak neutral hadronic current $j_z^\mu$.  The vector and axial vector
coupling constants, $C_{Vi}$ and $C_{Ai}$  are listed in Table 1. Numerical
values of the parameters that best fit data on  charged current semi-leptonic
decays of hyperons are (Gaillard \& Sauvage 1984):  D=0.756 , F=0.477,
$\sin^2\theta_W$=0.23 and $\sin\theta_c = 0.231$.  
Due to the large vector couplings of the $\Sigma^-$ and $\Xi^-$, $|C_V| \sim
2$, these particles, if present in protoneutron star matter, will give
significant contributions to neutrino scattering.  Neutrino scattering off
leptons in the same family involves  charged current couplings as well, and one
has to sum over both the  contributing diagrams.  At tree level, however, one
can express the total  coupling by means of a Fierz transformation; this is
accounted for in Table 1.

\vskip 10pt
Given the general structure of the neutrino coupling to matter, 
the differential cross section for elastic scattering for incoming neutrino
energy $E_{\nu}$ and outgoing neutrino energy $E_{\nu}^{'}$ is given by (Fetter
and Walecka 1971)
\begin{eqnarray}
\frac{1}{V}\frac{d^3\sigma}{d\Omega^{'2} dE_{\nu}^{'}}& =& -\frac{G^2}{128 
\pi^2}\frac{E_{\nu}^{'}}{E_{\nu}}~{\rm Im}~~(L_{\alpha\beta}\Pi^{\alpha\beta})
\,, 
\label{dcross}
\end{eqnarray}
where the neutrino tensor $L_{\alpha\beta}$ and the target particle 
polarization $\Pi^{\alpha\beta}$ are 
\begin{eqnarray}
L_{\alpha\beta} &=& 8[2k_{\alpha}k_{\beta}+(k\cdot q)g_{\alpha\beta}
-(k_{\alpha}q_{\beta}+q_{\alpha}k_{\beta})\mp i\epsilon_{\alpha\beta\mu\nu}
k^{\mu}q^{\nu}] \,, \\
\Pi_{\alpha\beta}^i & = & -i \int \frac{d^4p}{(2\pi)^4} \, 
{\rm Tr}~[G^i(p)J_{\alpha} G^i(p+q)J_{\beta}]\,. 
\end{eqnarray}
Above, $k_{\mu}$ is the incoming neutrino four momentum and $q_{\mu}$ is  the
four momentum  transfer.    The Greens' functions $G^i(p)$ (the index $i$
labels particle species)  depend on the Fermi momentum $k_{Fi}$ of target
particles.  In the Hartree approximation, the  propagators are obtained by
replacing $M_i$ and $k_{Fi}$  in the free particle propagators by $M^*_i$ and
$k^*_{Fi}$ (see below), respectively.  The current operator $J_{\mu}$ is
$\gamma_{\mu}$ for the vector current and $\gamma_{\mu}\gamma_5$ for the axial
current. Given the V-A structure of the particle currents, we have
\begin{eqnarray}
\Pi_{\alpha\beta}^i
&=&C_{Vi}^2\Pi_{\alpha\beta}^{V~i}+C_{Ai}^2\Pi_{\alpha\beta}^{A~i}
-2C_{Vi}C_{Ai}\Pi_{\alpha\beta}^{VA~i} \,.
\end{eqnarray}
For the vector polarization, 
$\{J_\alpha,J_\beta\}::\{\gamma_\alpha,\gamma_\beta\}$, 
for the axial
polarization,  $\{J_\alpha,J_\beta\} :: 
\{\gamma_\alpha\gamma_5,\gamma_\beta\gamma_5\}$ and for the mixed
part,  $\{J_\alpha,J_\beta\} ::\{\gamma_\alpha\gamma_5,
\gamma_\beta\}$.  Further, the polarizations contain two functions, 
the density dependent part that describes particle-hole excitations and 
the Feynman part that describes particle-antiparticle excitations.  For elastic
scattering,  with $q_{\mu}^2<0$,   the contribution of the  Feynman parts
vanish.   Using vector current conservation and translational invariance, 
$\Pi_{\alpha\beta}^V$ may be written in terms of two independent components. 
In a frame where $q_{\mu}=(q_0,|q|,0,0)$, we have  
\begin{eqnarray*}
\Pi_T = \Pi^V_{22} \qquad {\rm and} \qquad
\Pi_L = -\frac{q_{\mu}^2}{|q|^2}\Pi^V_{00} \,.
\end{eqnarray*}
The axial current-current correlation function can be written as a vector
piece plus a correction term: 
\begin{eqnarray}
\Pi_{\mu \nu}^A=\Pi^V_{\mu \nu}+g_{\mu \nu}\Pi^A \,.
\end{eqnarray}
The mixed, axial current-vector current correlation function is 
\begin{eqnarray}
\Pi_{\mu \nu}^{VA}= i\epsilon_{\mu, \nu,\alpha,0}q^{\alpha}\Pi^{VA}\,. 
\end{eqnarray}
The above mean field or Hartree polarizations, which characterize the
medium response to the neutrino, have been explicitly evaluated in previous
work (Horowitz \& Wehrberger 1991).  In terms of these polarizations, the 
differential cross-section is 
\begin{eqnarray}
\frac{1}{V}\frac{d^3\sigma}{d\Omega^{'2} dE_{\nu}^{'}}& =& -\frac{G^2}{16
\pi^3}~ \frac{E_{\nu}^{'}}{E_{\nu}} q_{\mu}^2 ~[AR_1+R_2+BR_3] 
\label{dcross2}
\end{eqnarray}
with
\begin{eqnarray}
A=\frac{2k_0(k_0-q_0)+q_{\mu}^2/2}{|q|^2}~~;
~~B=2k_0-q_0\,.
\end{eqnarray}
The polarizations may be combined into three uncorrelated response functions
$R_1,R_2$ and $R_3$ by summing over the contributions from each particle
species $i$: 
\begin{eqnarray}
R_1&=&\sum_{i}~[C_{Vi}^2+C_{Ai}^2]~[{\rm Im}~\Pi_T^i + {\rm Im}~\Pi_L^i],\\
\label{r1}
R_2&=&\sum_{i}~ C_{Vi}^2 ~{\rm Im}~\Pi_T^i + C_{Ai}^2 
~[{\rm Im}~\Pi_T^i - {\rm Im}~\Pi_A^i],\\
\label{r2}
R_3&=&\pm \sum_{i}~ 2 C_{Ai}C_{Ai}~{\rm Im}~\Pi_{VA}^i\,.
\label{r3}
\end{eqnarray}

The imaginary parts of the lowest order one-loop polarization parts,  required
for the differential cross-sections, are evaluated at finite density and
temperature.  These retarded polarizations characterize the medium response at
the mean field level.   These functions depend upon the individual
concentrations, which are  controlled by the effective chemical potentials
$\nu_i$ and the corresponding effective masses $M_i^*$, for which a many-body
description of the multi-component system is required (see \S 3). 

Analytic closed form expressions for the density dependent polarization
functions at zero temperature have been evaluated by Lim and
Horowitz (1989).  The finite temperature polarizations have been
investigated by Saito et al. (1989).  However, we are not aware
that closed form analytical formulae for the finite temperature retarded
polarizations have been  been reported elsewhere in the literature.  Here, we
provide simple analytical formulae for these polarizations in the space like
region ($q_0\le|q|$ and $q_{\mu}^2\le 0$).  The integral forms of the finite 
temperature correlation functions and their simplification are discussed in 
Appendix A.  The results for the various polarizations  (for one species of the
target particles) are:
\begin{eqnarray}
{\rm Im}~\Pi_T &=& \frac{q_{\mu}^2}{4 \pi q^3}[I_2+q_0I_1] 
+ \frac{1}{4 \pi q}
\left[\left(M^{*2}+\frac{q_{\mu}^2}{2}
+\frac{q_{\mu}^4}{4q^2}\right)I_0\right] \,,\\
{\rm Im}~ \Pi_L &=& \frac{q_{\mu}^2}{2 \pi q^3}
\left[I_2+q_0I_1+\frac{q_{\mu}^2}{4}I_0\right] \,,\\
{\rm Im}~\Pi_A &=& \frac{M^{*2}}{2 \pi q}I_0  \,,\\
{\rm Im}~\Pi_{VA}&=&\frac{q_{\mu}^2}{8 \pi q^3} [q_0 I_0+2I_1]\,.
\end{eqnarray}
Above, $I_0,I_1$, and $I_2$ are phase space integrals. 
Each phase space integral consists of two contributions: 
\begin{equation}
I_i=I_i^+ + I_i^- , ~~i=0,1~~{\rm and}~~2 \,,
\end{equation}
where the superscript $+$ refers to particle excitations and the superscript
$-$  to antiparticle excitations.  At finite temperature, both
particle and antiparticle excitations contribute, although 
the latter contribution is  exponentially suppressed by the factor $ \exp(-\mu
/T)$.  Explicit analytical forms for the phase space integrals are: 
\begin{eqnarray}
I_0^{\pm} &=& q_0-T\xi_1^{\pm}\,, \\
I_1^{\pm} &=& \pm \nu q_0 - \frac{q_0^2}{2}-T^2\xi_2^{\pm}-e_ 
- T\xi_1^{\pm} \,,\\
I_2^{\pm} &=& q_0^2 \nu^2 \mp \nu q_0^2 + 
\frac{\pi^2}{3}q_0T^2+\frac{q_0^3}{3} \nonumber \\
&+& 2T^3\xi_3^{\pm}-2e_-T^2\xi_2^{\pm}+e_-^2T\xi_1^{\pm} \,,
\label{ins}
\end{eqnarray} 
where the factors $\xi_n^{\pm}$ may be expressed as differences of 
polylogarithmic functions $Li_n$ as 
\begin{eqnarray}
\xi_n^{\pm} &=& Li_n(-\alpha_1^{\pm}) - Li_n(-\alpha_2^{\pm})\,; \\
\alpha_1^{\pm} &=& \exp\left(\frac{q_0+e_- \mp \nu}{kT}\right) 
\qquad {\rm and} \qquad 
\alpha_2^{\pm} = \exp\left(\frac{e_- \mp \nu}{kT}\right) \,.
\label{dlis}
\end{eqnarray}
The polylogarithimic functions 
\begin{equation}
Li_n(z) = \int_0^z \frac {Li_{n-1}(x)}{x} \,dx \,, 
\qquad Li_1(x) = \log (1-x)
\end{equation}
are defined to conform to the definitions of Lewin (1983). 

\noindent The factor $e_-$ in the above equations is the kinematic lower
limit and  is given by
\begin{equation}
e_-=-\frac{q_0}{2}+\frac{q}{2}\sqrt{1+4\frac{M^{*2}}{|q_{\mu}^2|}} \,\,.
\end{equation}
The quantities $M^*$ and $\nu$ are the effective mass and 
effective chemical potential of the particle, respectively (see \S 3). 

The above formulae are valid for $q_0\ge 0$.  We can extend them to the case 
$q_0\le 0$ using the principle of detailed balance, which ensures that 
\begin{equation}
{\rm Im}~\Pi(q_0\le 0) = \exp(-|q_0|/T)~{\rm Im}~\Pi(q_0\ge 0)  \,.
\end{equation}

These analytical expressions allow us to understand the behavior of the
correlation functions with both density and temperature.   This particular
representation for the  relativistic response functions permits one to
calculate accurately the  cross sections in all regimes of matter degeneracy. 
This is especially important in multi-component dense matter, where  one
expects to encounter different levels of degeneracy for each  particle species. 
Note also that this approach incorporates the effects of strong interactions
through modifications of the particle propagators.  In addition, the time
consuming and often problematic numerical integration required to evaluate the
phase space integrals (see Eq.~(\ref{pspace}) in Appendix A) is avoided.    

The total inclusive scattering rate for neutrinos from a hot and dense system
is obtained  by integrating over the allowed kinematic region in the
$\omega-|q|$ space. Explicitly, 
\begin{equation}
\frac{\sigma (E)}{V}=\frac{G^2}{2\pi^2 E^2}
\int^E_{- \infty}dq_0 
\int^{2E-q_0}_{|q_0|} d|q|~ |q|q_{\mu}^2~[AR_1+R_2+BR_3]\,.
\end{equation}
Note that the total cross-section per unit volume, which has the dimension of
inverse length, gives the inverse  collision mean free path.  Various other
transport coefficients, such as the diffusion coefficient and the thermal
conductivity, can also be obtained in a similar fashion.

\section{ COMPOSITION OF NEUTRON STAR MATTER }

To explore the influence of the presence of hyperons in dense matter, we 
employ a relativistic field theoretical model in which the interactions 
between baryons are mediated by the exchange of $\sigma,\omega$, and $\rho$ 
mesons.  The full Lagrangian density is given by~ (Serot \& Walecka 1986), 
\begin{eqnarray*}
L &=& L_H +L_\ell \nonumber \\
  &=& \sum_{B} \overline{B}(-i\gamma^{\mu}\partial_{\mu}-g_{\omega B}
\gamma^{\mu}
-g_{\rho B}\gamma^{\mu}{\bf{b}}_{\mu}\cdot{\bf t}-M_B+g_{\sigma B}\sigma)B \\ 
&-& \frac{1}{4}W_{\mu\nu}W^{\mu\nu}+\frac{1}{2}m_{\omega}^2\omega_{\mu}\omega^
{\mu} - \frac{1}{4}{\bf B_{\mu\nu}}{\bf 
B^{\mu\nu}}+\frac{1}{2}m_{\rho}^2 \rho_{\mu}\rho^{\mu}\\
&+& \frac{1}{2}\partial_{\mu}\sigma\partial^{\mu}\sigma +\frac{1}{2}
m_{\sigma}^2\sigma^2-U(\sigma)\\
&+& \sum_{l}\overline{l}(-i\gamma^{\mu}\partial_{\mu}-m_l)l \,.
\end{eqnarray*}
Here, $B$ are the Dirac spinors for baryons and $\bf t$ is the isospin
operator. The sums include baryons $B=n,p,\Lambda,\Sigma$, and $\Xi$, and 
leptons, $l = e^-$ and $\mu^-$. The field strength tensors for the $\omega$ 
and 
$\rho$ mesons are $W_{\mu\nu} = \partial_\mu\omega_\nu-\partial_\nu\omega_\mu$
and ${\bf B}_{\mu\nu} =  \partial_\mu{\bf b}_\nu-\partial_\nu{\bf b}_\mu$ ,
respectively.  The potential $U(\sigma)$ represents the self-interactions of
the scalar field and is taken to be of the form 
\begin{eqnarray}
U(\sigma) =  \frac{1}{3}bM_n(g_{\sigma N}\sigma)^3 + \frac{1}{4}c(g_{\sigma 
N}\sigma)^4\,.
\end{eqnarray} 
Electrons and muons are included in the model as non-interacting particles,
since their interactions give small contributions compared to those of 
their free Fermi gas parts. 

In the mean field approximation, the partition function (denoted by $Z_H$) for
the hadronic degrees of freedom is given by
\begin{eqnarray}
\ln Z_H&\pj=&\pj\beta V\left[\thalf m_{\omega}^2\omega_0^2+\thalf 
m_{\rho}^2b_0^2 - \thalf m_{\sigma}^2\sigma^2-U(\sigma)\right]\nonumber\\
&&\qquad\qquad+ 2V\sum_B \int\frac{d^3k}{(2\pi)^3} \,\ln\left(1+{\rm e}
^{-\beta(E^*_B-\nu_B)}\right)\,,
\label{hyp2}
\end{eqnarray}
where $\beta = (kT)^{-1}$ and $V$ is the volume. 
The contribution of antibaryons is
not significant for the thermodynamics of interest here, and is therefore
not included in Eq.~(\ref{hyp2}). Here, the effective baryon masses 
$M_B^*=M_B-g_{\sigma B}\sigma$ and
$E^*_B=\sqrt{k^2+M^{*2}_B}$. The chemical potentials are given by
\begin{equation}
\mu_B=\nu_B+g_{\omega B}\omega_0+g_{\rho B}t_{3B}b_0\;,\label{hyp3}
\end{equation}
where $t_{3B}$ is the third component of isospin for the baryon. Note
that particles with $t_{3B}=0$, such as the $\Lambda$ and $\Sigma^0$   
do not couple to the $\rho$.  The effective chemical potential 
$\nu_B$ sets the scale of the temperature dependence of the thermodynamical
functions. 

Using $Z_H$, the thermodynamic quantities can be obtained in the standard way. 
The pressure $P_H=TV^{-1}\ln Z_H$, the number density for species $B$, and the
energy density $\varepsilon_H$ are given by
\begin{eqnarray}
n_B&\pj=&\pj2\int\frac{d^3k}{(2\pi)^3} 
\left({\rm e}^{\beta(E^*_B-\nu_B)}+1\right)^{\!-1}\;,\nonumber\\
\varepsilon_H&\pj=&\pj\thalf m_{\sigma}^2\sigma^2+U(\sigma)+
\thalf m_{\omega}^2\omega_0^2+\thalf m_{\rho}^2 b_0^2 
+2\sum_B
\int\frac{d^3k}{(2\pi)^3} 
E_B^*\left({\rm e}^{\beta(E^*_B-\nu_B)}+1\right)^{\!-1}\;.\label{hyp4}
\end{eqnarray}
The entropy density is then given by
$s_H=\beta(\varepsilon_H+P_H-\sum_B\mu_Bn_B)$.

The meson fields are obtained by extremization of the partition function, 
which yields the equations
\begin{eqnarray}
&&\fpj m_{\omega}^2\omega_0=\sum_Bg_{\omega B} n_B\quad;\quad
m_{\rho}^2b_0=\sum_Bg_{\rho B}t_{3B}n_B\;,\nonumber\\
&&\fpj m_{\sigma}^2\sigma=-\frac{dU(\sigma)}{d\sigma}
+\sum_B g_{\sigma B} ~~2\hspace{-1mm}\int\!\frac{d^3k}{(2\pi)^3} 
\frac{M_B^*}{E_B^*}\left({\rm e}^{\beta(E^*_B-\nu_B)}+1\right)^{\!-1} \,.
\label{hyp5}
\end{eqnarray}
The total partition function $Z_{total}=Z_HZ_L$, where $Z_L$ is the standard
non-interacting partition function of the leptons. 

The additional conditions needed to obtain a solution are provided by the
charge neutrality requirement, and, when neutrinos are not 
trapped, the set of equilibrium chemical 
potential relations required by the general condition
\begin{eqnarray}
\mu_i = b_i\mu_n - q_i\mu_l\,, 
\label{beta}
\end{eqnarray}
where $b_i$ is the baryon number of particle $i$ and $q_i$ is its
charge.  
For example, when $\ell=e^-$, this implies the equalities  
\begin{eqnarray}
&&\fpj \mu_{\Lambda} = \mu_{\Sigma^0} = \mu_{\Xi^0} = \mu_n \,, \nonumber \\
&&\fpj \mu_{\Sigma^-} = \mu_{\Xi^-} = \mu_n+\mu_e \,, \nonumber \\
&&\fpj \mu_p = \mu_{\Sigma^+} = \mu_n - \mu_e \,. 
\label{murel}
\end{eqnarray}
In the case that the neutrinos are trapped, Eq. (\ref{beta}) is replaced by
\begin{eqnarray}
\mu_i = b_i\mu_n - q_i(\mu_l-\mu_{\nu_\ell})\,.
\label{tbeta}
\end{eqnarray}
The new equalities are then obtained by the replacement
$\mu_e \rightarrow \mu_e - \mu_{\nu_e}$ in Eq.~(\ref{murel}).  The
introduction of additional variables, the neutrino  chemical potentials,
requires additional constraints, which we supply by fixing the lepton fractions,
$Y_{L\ell}$, appropriate for conditions prevailing in the evolution of the
protoneutron star.  The contribution to pressure from neutrinos of a given
species is $P_\nu=(1/24\pi^2)\mu_\nu^4$. 

In the nucleon sector, the constants $ g_{\sigma N}, g_{\omega N}, 
g_{\rho N}, b$, and $c$ are determined by reproducing the nuclear matter 
equilibrium density $n_0=0.16~{\rm fm}^{-3}$, and the binding energy per
nucleon ($\sim 16$ MeV),  the symmetry energy ($\sim 30-35$ MeV), 
the compression modulus ($200~{\rm MeV} \leq K_0 \leq 300~{\rm MeV})$, and the
nucleon Dirac effective mass $M^*=(0.6-0.7)~\times 939$ MeV at $n_0$. 
Numerical values of the coupling constants so chosen are shown in Table 2. This
particular choice of model  parameters are from Glendenning and Moszkowski
(1991) and will be referred to as GM1 hereafter. The prevalent uncertainty in
the  nuclear matter compression modulus  and the effective mass $M^*$ does not
allow for a unique choice of these  coupling constants.   The high density
behaviour of  the EOS  is sensitive to the strength of the meson coupling
constants employed.  Lacking definitive experimental and theoretical
constraints, this choice of parameters may be considered typical.       

The hyperon coupling constants may be determined by
reproducing the binding energy of the $\Lambda$ hyperon in nuclear matter
(Glendenning \& Moszkowski 1991). 
Parametrizing the hyperon-meson couplings in terms of nucleon-meson couplings 
through  
\begin{eqnarray} 
x_{\sigma H}=g_{\sigma H}/g_{\sigma N},~~~
x_{\omega H}=g_{\omega H}/g_{\omega N}
,~~~x_{\rho H}=g_{\rho H}/g_{\rho N} \,, 
\end{eqnarray}
the $\Lambda$ binding energy at nuclear density is given by 
\begin{eqnarray} 
(B/A)_\Lambda = -28 = x_{\omega \Lambda} g_{\omega N} \omega_0 
- x_{\sigma \Lambda} g_{\sigma N} \sigma_0\,, 
\end{eqnarray}
in units of MeV.  Thus, a particular choice of  $x_{\sigma
\Lambda}$ determines $x_{\omega \Lambda}$ uniquely.   To keep the number of
parameters small, the coupling constant
ratios for all the different hyperons are assumed to be the same. 
That is             
\begin{eqnarray} 
x_\sigma = x_{\sigma\Lambda} = x_{\sigma\Sigma} = x_{\sigma\Xi} = 0.6\,,
\end{eqnarray}
and similarly for the $\omega$ 
\begin{eqnarray} 
x_\omega = x_{\omega\Lambda} = x_{\omega\Sigma} = x_{\omega\Xi} = 0.653 \,.
\end{eqnarray}
The $\rho$-coupling is of
less consequence and is taken to be of similar order, i.e. 
$x_\rho = x_\sigma$\,.

\subsection{Composition in a Cold Catalyzed Neutron Star }

\begin{figure}
\begin{center}
\leavevmode
\epsfxsize=6.0in
\epsfysize=6.0in
\epsffile{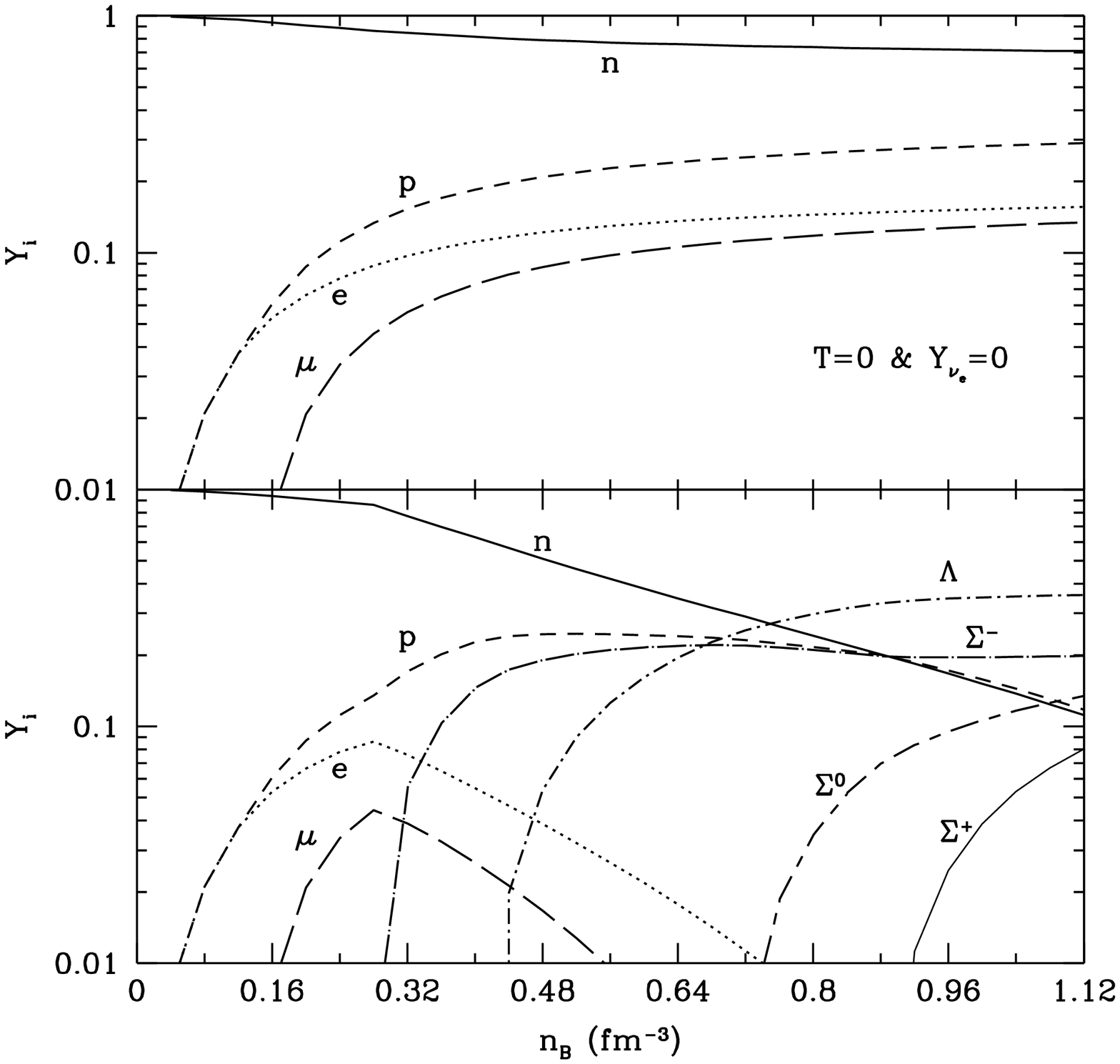}
\end{center}
\caption[]{\footnotesize }
{\label{pfracs}}
\end{figure}

Old and cold neutron stars are essentially neutrino free. 
In Figure~\ref{pfracs}, we show the relative fractions, $Y_i=n_i/n_b$, 
of the baryons and
leptons in charge neutral and $\beta-$equilibrated neutrino-free matter  at
zero temperature.  The upper panel shows the concentrations for the case in
which only the nucleonic degrees of freedom are allowed.  The lower panel
contains results for the case in which hyperonic degrees of freedom are also
allowed.  For the model parameters chosen, the $\Sigma^-$ hyperon appears at a
density lower than the $\Lambda$ hyperon.  This is because the somewhat higher
mass of the $\Sigma^-$ is compensated by the presence of the $e^-$ chemical
potential in the equilibrium condition of the $\Sigma^-$. More massive and
more positively charged particles appear at higher densities. With the
appearance of the negatively charged $\Sigma^-$, which competes with leptons in
maintaining charge neutrality, the lepton concentrations begin to fall.  The
important point is that, with increasing density, the system contains many
baryon species with nearly equal concentrations.                    
                                 
It should be pointed out, however, that moderate changes in the poorly known 
$\Sigma$ and $\Xi$ couplings has large effects on the appearance of negatively
charged particles (Prakash et al., 1996).   Increasing the coupling constants
of a hyperon species delays its appearance to a higher density.  This is
because  the threshold condition, Eq.~(\ref{tbeta}), receives contributions
from the $\sigma, \omega$, and $\rho$ fields, the net result being positive due
to that of the $\omega$. If all the couplings are scaled up, the positive
contribution becomes larger, and hence the appearance of the particle is
delayed to a higher density.  Clearly, the thresholds for the strange particles
are sensitive to the coupling constants, which are presently poorly 
constrained by either theory or experiment.  Notwithstanding these caveats, it
is clear that one or the other hyperon species is likely to exist in dense 
matter.

\subsection{Composition in an Evolving Newly Born Neutron Star}

The protoneutron star formed subsequent to the core bounce, and its early 
evolution has been investigated in earlier works (Burrows \& Lattimer 1986,
Keil \& Janka 1995).  Detailed studies of the 
dynamics of core collapse and supernova indicate that within 
milliseconds of the shock wave formation the core settles into hydrostatic 
equilibrium, with a low entropy and large lepton content.  The electron  lepton
fraction  $Y_{Le}=Y_{e}+Y_{\nu_e}$ at bounce in the interior is estimated to be
about  $0.4$. The electron type neutrinos formed and trapped in the core 
during collapse are degenerate with a chemical potential of about 300 MeV.  In
the case of muons it is generally true that, unless $\mu > m_\mu c^2$, the net
number of $\mu$'s or $\nu_\mu$'s present is zero. Because no muon-flavor
leptons are present at the onset of trapping, $Y_{\nu_\mu}=-Y_\mu$. Following
deleptonization, $Y_{\nu_\mu}=0$, and $Y_\mu$ is determined by $\mu_\mu=\mu_e$
for $\mu_e > m_\mu c^2$ and is zero otherwise.  

The entropy per baryon, $S$, in the interior is low, of order unity, which 
corresponds to temperatures between  5-30 MeV in the interior.  The temperature
and entropy increases as one moves  from the center outwards (Burrows \&
Lattimer 1986). At the very early stage, the electron neutrino chemical
potential  is largest in the center and drops appreciably as a function of the 
distance from the center. This gradient in the electron neutrino chemical
potential is primarily  responsible for driving the deleptonization phase. 
During the early  deleptonization phase, the neutrino  re-heating of matter
increases the central temperature. This, coupled with  other diffusive
processes, reverses the temperature gradient. On time scales of about 10-15 s,
the central temperature is raised to  about 50 MeV  ($S \sim 2$) and decreases
outwards.  This marks the onset of the cooling phase in the central regions of
the star.  The conditions at the onset of deleptonization and cooling are thus 
significantly different. In summary, the composition and temperature in the 
central regions of the star at the beginning of deleptonization are 
characterized by a high lepton fraction $(Y_{Le}=0.4)$ and low entropy  $(S\sim
1)$, while the cooling phase is characterized by a low neutrino fraction
$(Y_{\nu_e} \sim 0)$  and high entropy $(S \sim 2)$.    
                          
For a full treatment of neutrino  transport during the evolution,
opacities for a wide range of composition and matter degeneracy are required.
The dynamical changes in the lepton fraction and the temperature modify the
composition of matter and the typical neutrino energies in the inner core. In
addition, there are structural  changes in the interior associated with changes
in $Y_{Le}$ and $T$  (Prakash et al., 1996).  Computer simulations of the
evolution account for these effects dynamically. Here, we have identified two
important and distinct phases during the early evolution based on such
simulations (Burrows \& Lattimer 1986; Keil \& Janka 1995), in order to
highlight the role of neutrino scattering.

\subsubsection {The Deleptonization Phase }

\begin{figure}
\begin{center}
\leavevmode
\epsfxsize=6.0in
\epsfysize=6.0in
\epsffile{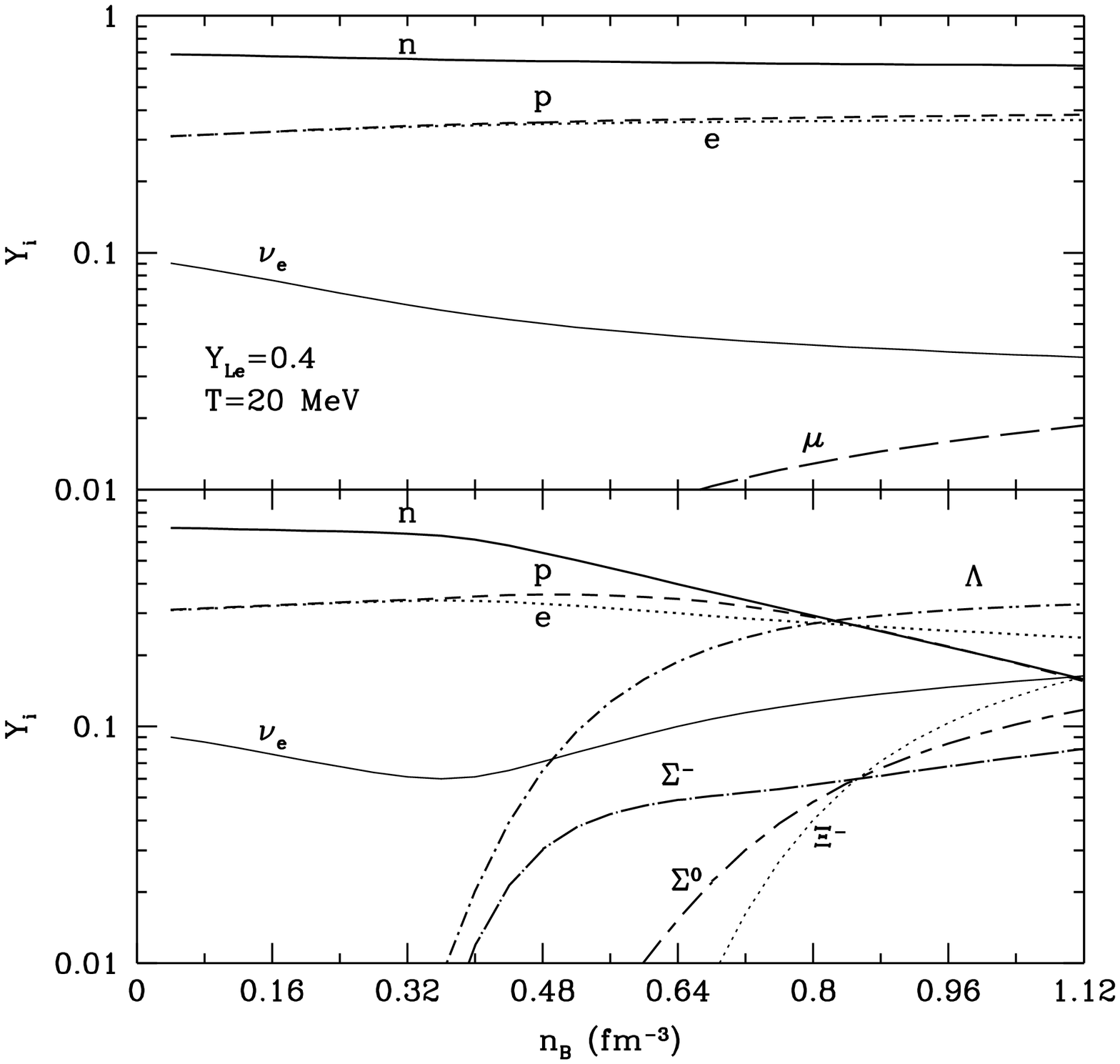}
\end{center}
\caption[]{\footnotesize }
{\label{pfracd}}
\end{figure}

For typical conditions in this phase,  $Y_{Le}=0.4$ and $T=20$ MeV,  Figure
\ref{pfracd} shows the concentrations of the various species. The  top panel
shows results for matter with nucleons only.  The bottom panel  refers to
the case in which hyperons are present.  In both cases, neutrino  trapping
significantly alters the composition from the neutrino-free case (see
Figure~\ref{pfracs}).   This is because $\mu=\mu_e-\mu_{\nu_e}$ is much smaller
than $\mu_e$ in the neutrino-free  case. This results in large electron, and
hence, to satisfy charge neutrality, large proton concentrations.  Also, the
appearance of hyperonic components is delayed to higher densities; in
particular, the concentrations of electrically charged hyperons  are
suppressed.   Finite temperature  effects on the composition are less
significant, and in general favors the presence  of strange baryons.   Relative
to nucleons-only matter,  the neutrino chemical potential and hence the
neutrino concentrations increase  substantially with density in matter
containing hyperons.  This introduces an important distinction between the
opacities in matter containing only nucleons and in matter containing hyperons
as well.

\subsubsection {The Cooling Phase }

\begin{figure}
\begin{center}
\leavevmode
\epsfxsize=6.0in
\epsfysize=6.0in
\epsffile{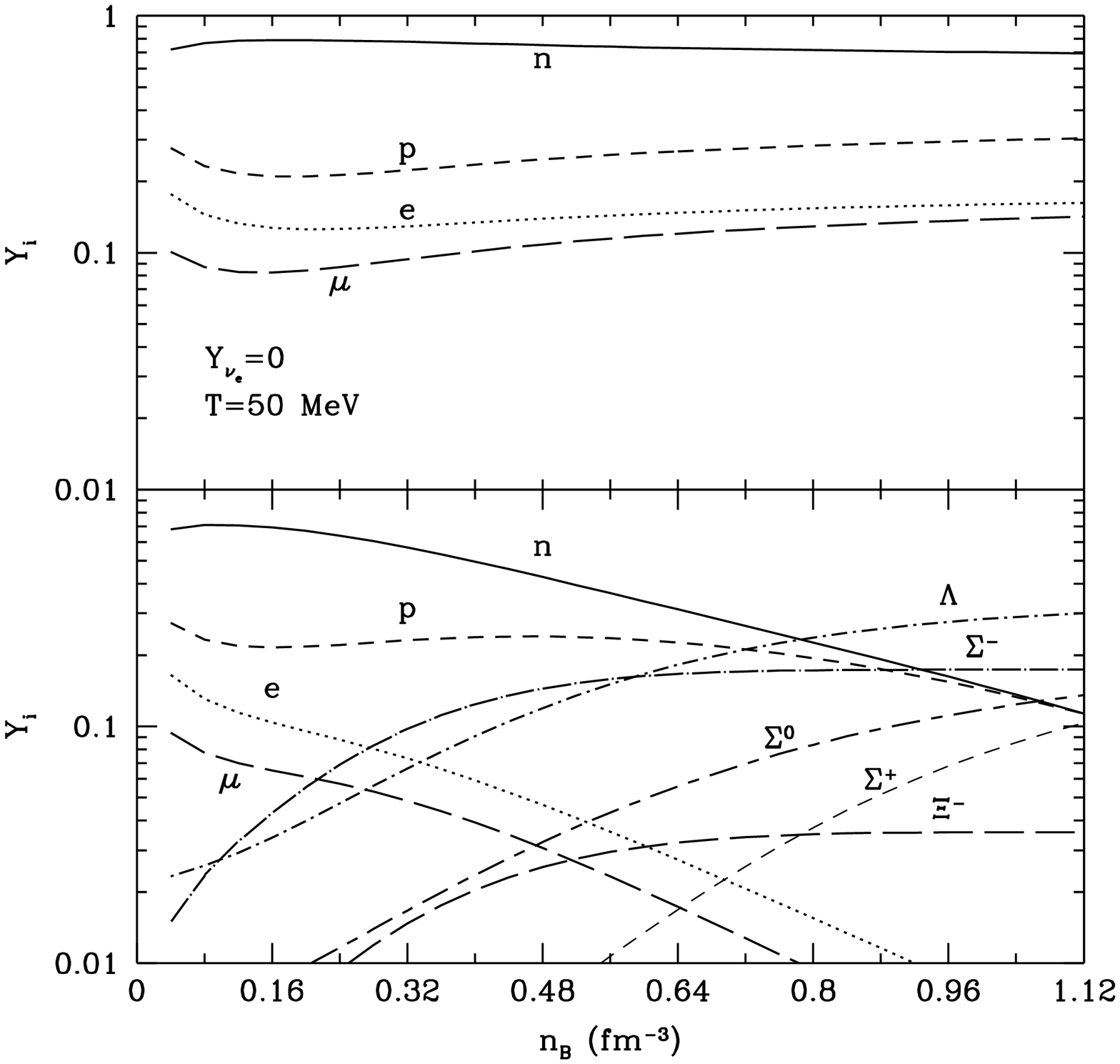}
\end{center}
\caption[]{\footnotesize }
{\label{pfracc}}
\end{figure}

In this phase, $\mu_{\nu_e} \simeq 0 $.   Figure \ref{pfracc} shows the 
concentrations $Y_i$ for a typical $ T=50$ MeV.   Here, the 
strangeness-bearing components occur at significantly lower densities  than in
the neutrino-trapped case.   A comparison of the two panels shows that in
strangeness-rich matter, the $\Sigma^-$ hyperons effectively replace the
leptons in maintaining charge neutrality.  The increasing abundance of the
neutral particles $\Lambda$ and $\Sigma^0$ 
has important consequences for neutrino scattering.  
Although the lowest
order contributions from the neutral $\Lambda$ and $\Sigma^0$ are smaller than
those of the $\Sigma^-$ and $\Xi^-$ (see Table 1),
their presence  furnishes baryon
number, which decreases the relative concentrations of nucleons.  (In contrast,
neutrons are the most abundant particles in nucleons-only matter.)  Relative to
neutrino trapped matter (Figure \ref{pfracd}), the larger temperature and the
absence of a  neutrino chemical potential both contribute synergetically to
enhance the hyperonic fractions.

There exists an unambiguous difference between  the deleptonization
(neutrino-trapped) and cooling (neutrino-free) phases  (Prakash et al., 1996). 
Neutrino trapping delays the appearance of hyperonic components to higher
baryon  densities.  This implies that during the early deleptonization phase, 
matter consists mostly of non-strange baryons, except possibly at high 
densities.  The  cooling phase is characterized by the presence of a
substantial amount of strangeness-rich hyperons, as they appear at lower
densities. Thus, we may expect modifications to the neutral current scattering
due to strangeness to be quantitatively different during the two phases.   In
the following section, we present the scattering cross sections in both these
phases.

\section {NEUTRINO SCATTERING CROSS SECTIONS }

We turn now to quantitative results for the neutrino scattering rates. The
differential and total cross sections are evaluated per unit volume of matter,
and the contribution from  each particle species is summed over. The results
presented here take into account strong interaction and finite temperature
effects through the  finite temperature mean field response functions.  
In order to calculate the differential cross sections   using Eq.
(\ref{dcross2}),  the chemical potentials of all particle species, the
temperature and the mean fields need to be specified.   These are provided by
the calculations described in \S 3.  For the EOS employed, the density in the
central regions of the star is in the range  of $3-7$ times the nuclear
saturation density (Prakash et al 1996), so for the most part, 
we choose to focus on a representative density of $n_B=0.64~ \rm fm^{-3}$.

\subsection {Influence of Composition and Temperature}

\begin{figure}
\begin{center}
\leavevmode
\epsfxsize=6.0in
\epsfysize=6.0in
\epsffile{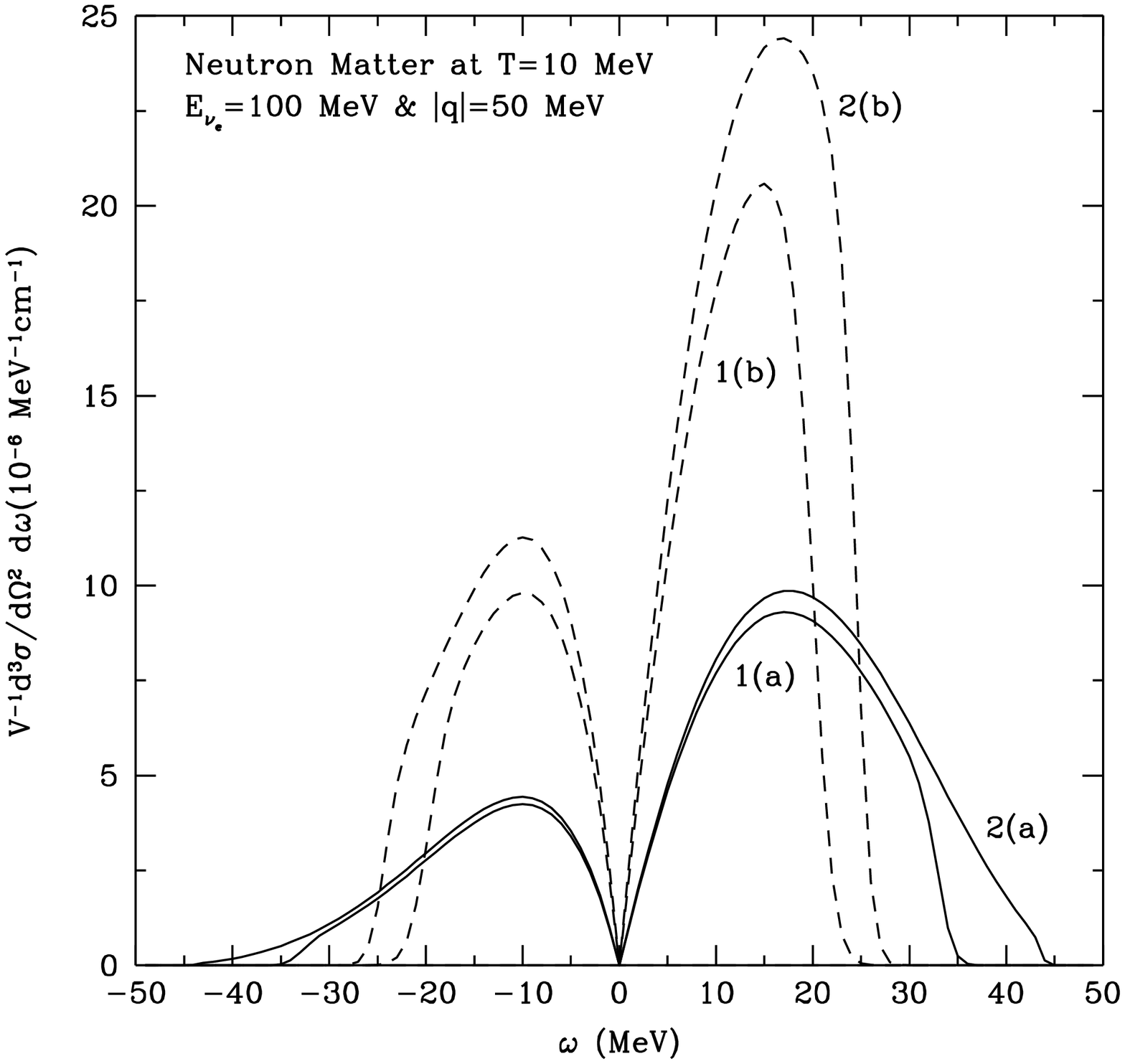}
\end{center}
\caption[]{\footnotesize }
{\label{rhods}}
\end{figure}

A first orientation to the  interaction corrections and
finite temperature  effects on the neutrino scattering rates is provided by
investigating pure neutron matter.   Figure \ref{rhods} shows the effect of
varying density on the differential cross-sections (solid curves 1(a) and
2(a)).   The curve labels 1 and 2 refer to $n_B=0.32~{\rm fm}^{-3}$ and 
$n_B=0.64~{\rm fm}^{-3}$, respectively.  For comparison, the 
corresponding  results
for a free Fermi gas are also shown (dashed curves 1(b) and 2(b)).   
Relativistic effects from  the matrix elements and phase space considerations
are both small for  non-interacting baryons.   These effects, however,
significantly influence the mean field response.   In particular, the response
is sensitive to  the effective mass and the effective chemical potential.  
Scalar interactions, which generate nucleon effective masses that decrease with
increasing density, considerably alter the response  in comparison to that of
a free Fermi gas.  For small $\omega$, interactions lead to a suppressed cross
section, while the large $\omega$ cross sections are enhanced.

\begin{figure}
\begin{center}
\leavevmode
\epsfxsize=6.0in
\epsfysize=6.0in
\epsffile{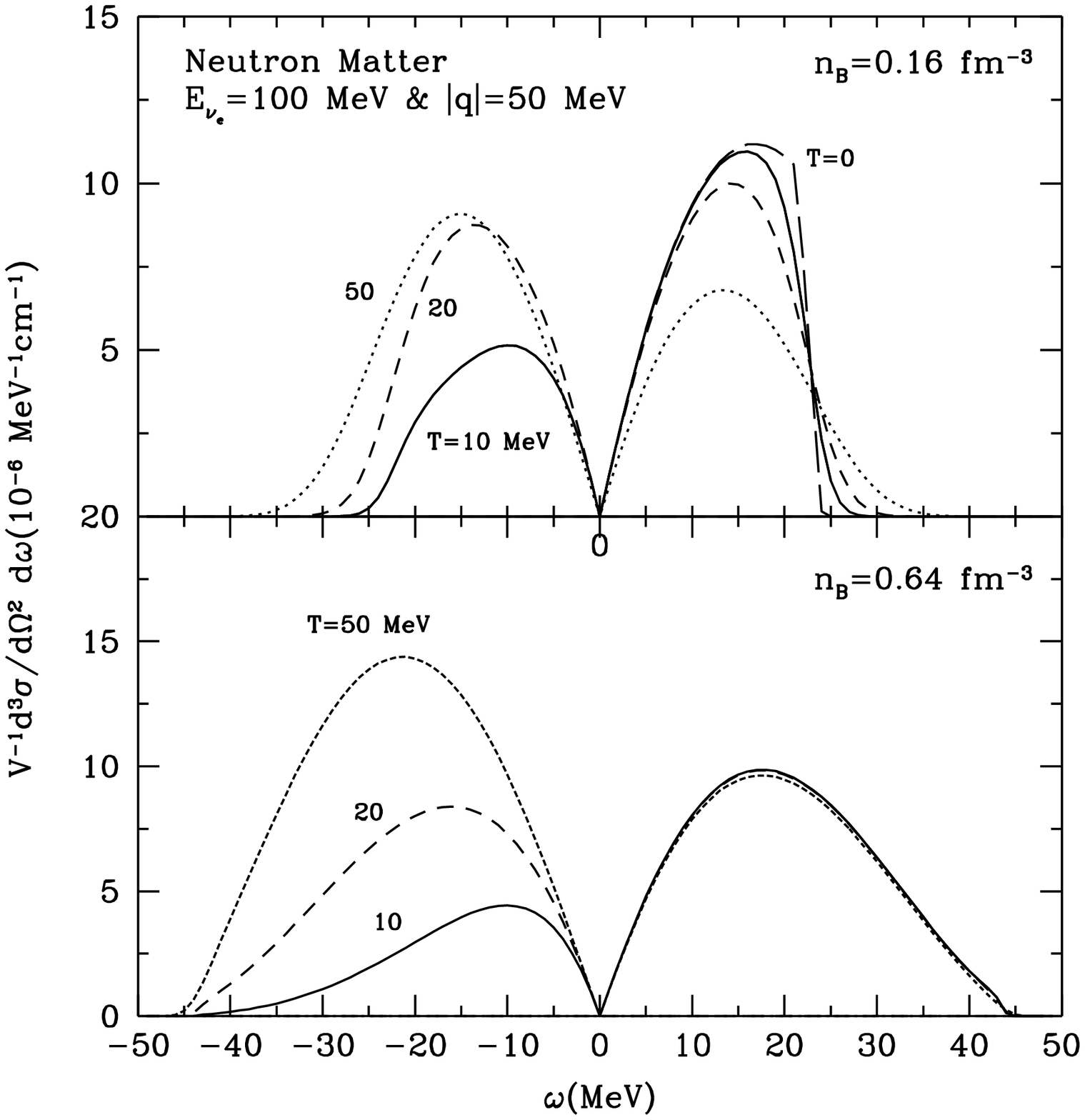}
\end{center}
\caption[]{\footnotesize }
{\label{tds}}
\end{figure}

Figure \ref{tds} illustrates finite temperature effects  on the differential
cross sections in pure neutron matter.  At zero temperature, only the positive
$\omega$ excitations are present, since negative $\omega$ excitations are
blocked by the Pauli principle.   However, changing the temperature on the
scale of the  energy transfer $\omega$ furnishes sufficient number of 
particles in   excited states.  This gives rise to a  significant amount of
negative $\omega$ response (Iwamoto \& Pethick 1982; Sawyer 1989).  This is
true even if the temperature is very small compared with the target  particle
Fermi energies.   For positive  $\omega$, finite temperature corrections depend
on the density.  For low baryon densities, even temperatures of order a few
tens of MeV  render the system non-degenerate. Hence, the positive $\omega$
response shows considerable sensitivity to temperature (upper panel).  With
increasing density, when matter becomes increasingly degenerate, effects of
temperature are negligibly small for $\omega >0$ (lower panel).

\begin{figure}
\begin{center}
\leavevmode
\epsfxsize=6.0in
\epsfysize=6.0in
\epsffile{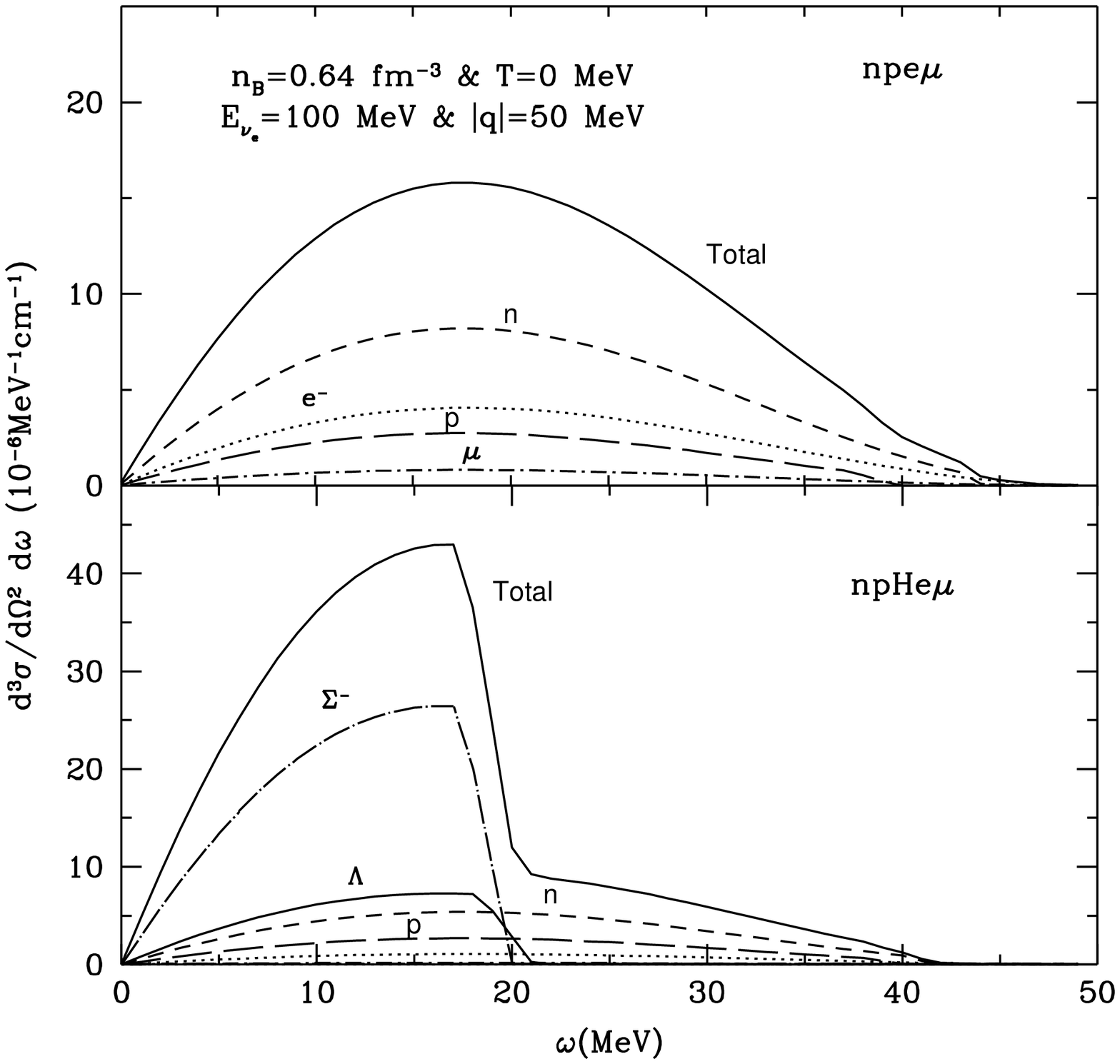}
\end{center}
\caption[]{\footnotesize }
{\label{compds1}}
\end{figure}

\begin{figure}
\begin{center}
\leavevmode
\epsfxsize=6.0in
\epsfysize=6.0in
\epsffile{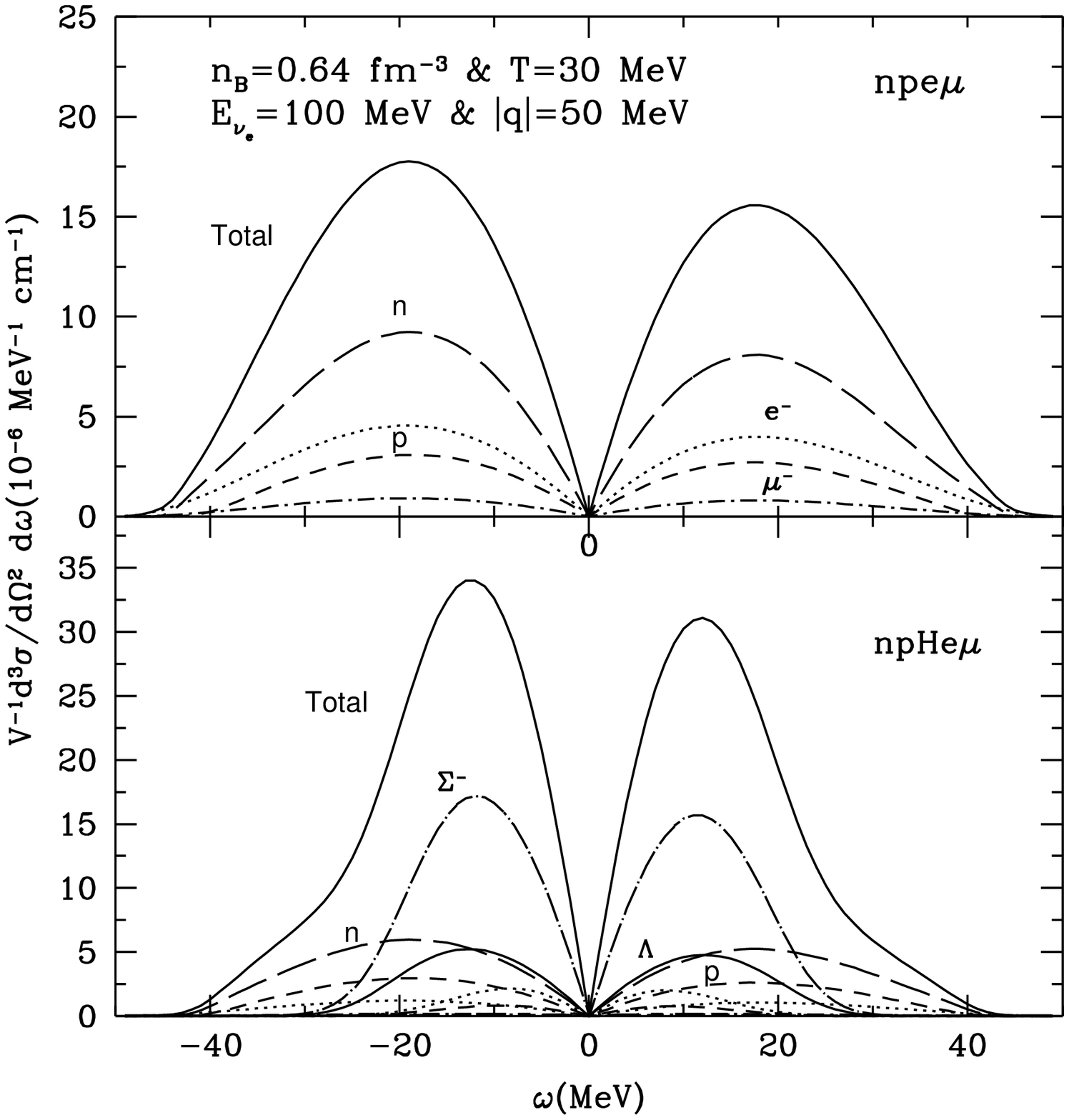}
\end{center}
\caption[]{\footnotesize }
{\label{compds2}}
\end{figure}

The response of a multi-component system is significantly different from that
of a single component system.  At a given density, the concentrations and the
effective masses of the individual particles, and, their specific coupling to the
neutrinos  determine the  total response.    These features are illustrated in
Figures \ref {compds1} and \ref{compds2}, where we  contrast the differential 
cross sections at  $n_B=0.64 ~\rm fm^{-3}$ and $T=30$ MeV in nucleonic matter
and in matter with strange baryons.   The  contribution of each particle
species is shown separately in these figures.  At zero temperature (Figure
\ref{compds1}),  the particles are all degenerate and only positive $\omega$
excitations are present.   The contributions from the  $\Sigma^-$ and the
$\Lambda$  dominate the neutron contribution in the  low $\omega$ region 
(see lower left panel).    This may be roughly understood in terms of the
magnitudes of the product of two factors for the various particles: The first
factor is $\nu_i^2 = k_{F_i}^2 + M_i^{*^2}$ (true only at $T=0$), which
reflects the composition and the effective mass.  The second factor is 
$C_{V_i}^2 + 3 C_{A_i}^2$ and is a measure of the neutrino coupling.  The
dominance of the $\Sigma^-$ is due to the fact that both factors above are much
larger than those for the  $\Lambda$ and the neutron.  Notice also that  since
$\nu_\Lambda > \nu_n$ at this density,  the contribution from the $\Lambda$s  is
comparable to those from neutrons despite the fact its effective 
coupling is relatively
smaller (see Table. 1).    At finite temperature (Figure \ref{compds2}), the
$\omega <0$ response is opened up. Further, the sharp fall off in the cross
section  (for $\omega >0$), associated with the $\Sigma^-$ hyperon, is smoothed 
out.  Due to the varying degeneracies, the response of the different particle
species are different.  Most significant changes occur in the region where 
$\omega/T \sim 1$.   

In what follows, we study the differential and total cross sections at  finite
temperatures,  densities and composition relevant to the early evolution of a
newly born neutron star.

\subsection {The Deleptonization Phase }

\begin{figure}
\begin{center}
\leavevmode
\epsfxsize=6.0in
\epsfysize=6.0in
\epsffile{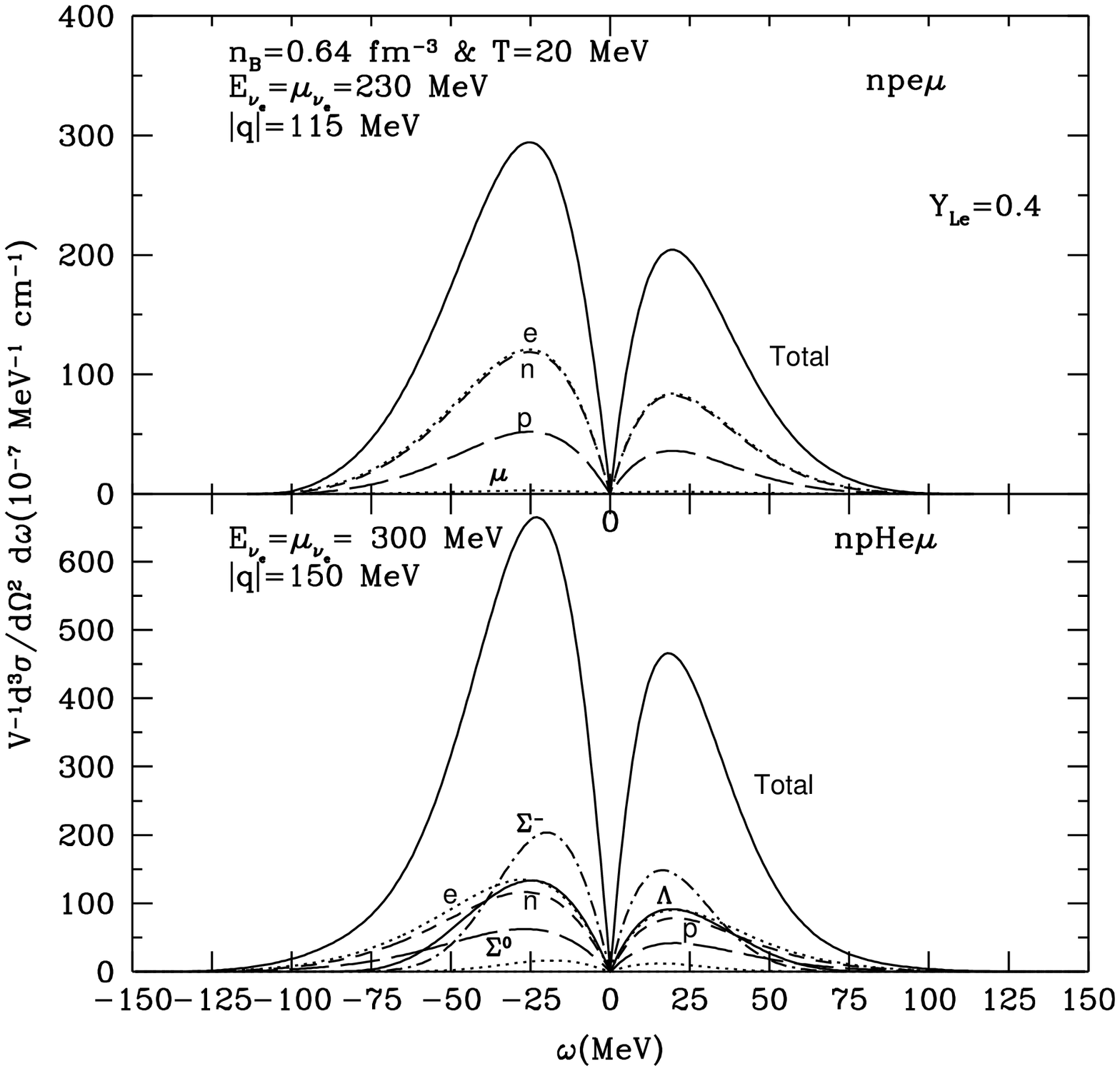}
\end{center}
\caption[]{\footnotesize }
{\label{dds}}
\end{figure}

Here, we calculate the cross sections  at a baryon number  density $n_B=0.64~
\rm fm^{-3}$, temperature  $T=20$ MeV and lepton fraction $Y_{Le}=0.4$, which
are representative  values in the central regions at the beginning of this
phase.  The  composition under these conditions was discussed in \S 3.2.1.  
Since neutrinos are degenerate, the relevant neutrino energies  lie in the
range  250--450 MeV.  The effects of neutrino degeneracy  are incorporated by
explicitly by including the neutrino final state blocking factor
$1-f(E-\mu_{\nu_e}+\omega)$ in the expression for the differential scattering
cross section.  The neutrino
chemical potential in the interior increases with density (since $Y_{Le}$ is
held constant at the value 0.4).  At a given  density,
only neutrinos close to the Fermi surface can actively participate in the 
diffusion process.  The different neutrino energies chosen reflect the
different  neutrino chemical potentials in matter with nucleons and hyperonic
components.  Figure \ref{dds} shows a comparison of the differential cross
sections in normal and strangeness-rich matter.  Note that electrons, due to
their large concentrations, contribute nearly as much to the cross sections as
the neutral particles.   Further, the $\Sigma^-$ hyperons provide the dominant
contributions.

\begin{figure}
\begin{center}
\leavevmode
\epsfxsize=6.0in
\epsfysize=6.0in
\epsffile{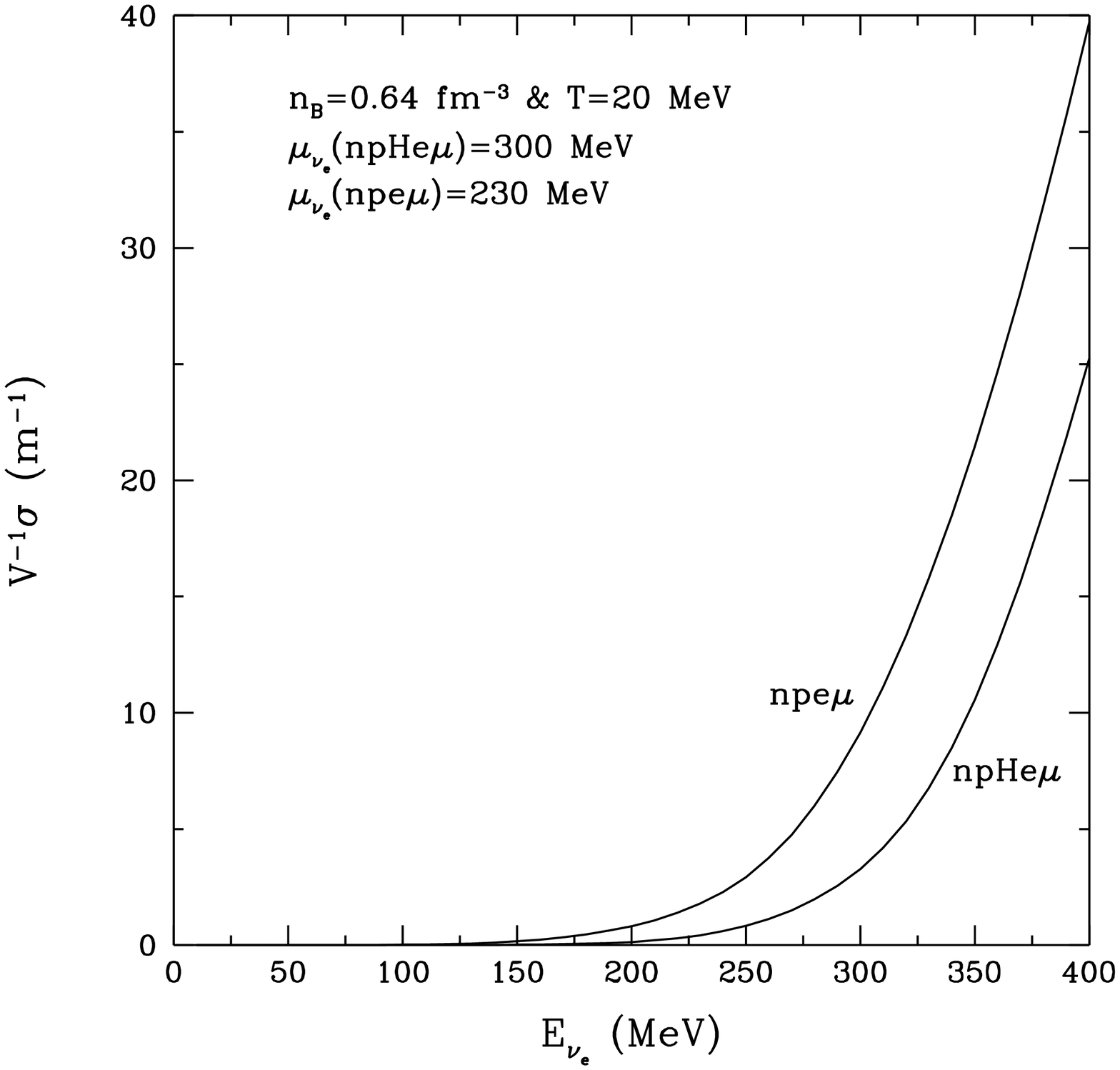}
\end{center}
\caption[]{\footnotesize }
{\label{sigede}}
\end{figure}

\begin{figure}
\begin{center}
\leavevmode
\epsfxsize=6.0in
\epsfysize=6.0in
\epsffile{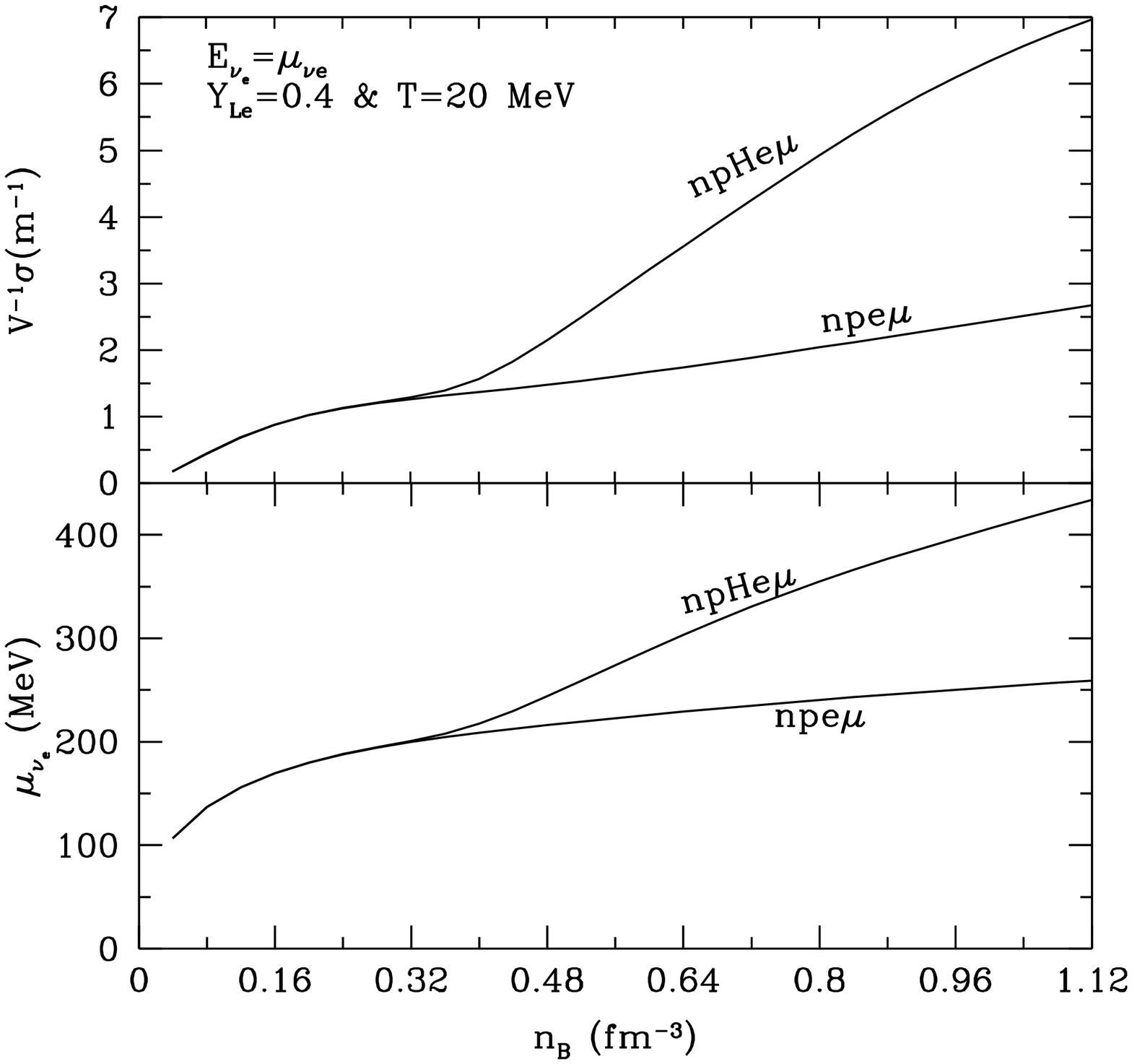}
\end{center}
\caption[]{\footnotesize }
{\label{sigedd}}
\end{figure}

The total cross sections are calculated by integrating over the allowed
kinematical region, accounting for the final state neutrino blocking.  
Figure \ref{sigede} shows the energy dependence of the cross sections at the
fiducial values of $n_B=0.64~{\rm fm}^{-3}$ and $T=20$ MeV.  The neutrino
chemical potentials in matter with and without hyperons are set by the
equilibrium conditions.  Neutrino final state blocking effectively suppresses
the cross sections for $E_{\nu_e} < \mu_{\nu_e}$.  Neutrinos of higher 
energy are unaffected by blocking; hence, the cross section grows rapidly.  
In the degenerate regime, however, lepton number and energy diffusion are
dominated by neutrinos close to the Fermi surface. (Initial state 
probabilities of high energy neutrinos are small.) Thus, an appropriate choice
for the neutrino energy is  the local neutrino chemical potential, i.e.
$E_{\nu_e}=\mu_{\nu_e}$.   In Figure \ref{sigedd}, we  show the density
dependence of the cross-sections (upper  panel) in normal and strangeness-rich
matter. The corresponding  neutrino chemical potentials are shown in the lower
panel.  The presence of strangeness causes the neutrino concentrations to be
significantly larger than those in normal matter.   As a result,  the cross
sections rise sharply with  density in the presence of strangeness,  chiefly due
to the substantial increase in the neutrino energy.

\subsection {The Cooling Phase }

Here, we present results for thermal electron type neutrinos. The
extension to  $\mu$ and $\tau$ type neutrinos and their anti-particles,   by
incorporating the appropriate neutrino-lepton coupling at tree level, is
straightforward.  (Neutrino coupling to leptons is flavour specific, 
but their coupling at tree level to the baryons is not.)
The transport of non-degenerate thermal  $\mu$ and $\tau$ neutrinos is 
important at all times, and thermal electron neutrino transport is important 
subsequent to deleptonization.  

\begin{figure}
\begin{center}
\leavevmode
\epsfxsize=6.0in
\epsfysize=6.0in
\epsffile{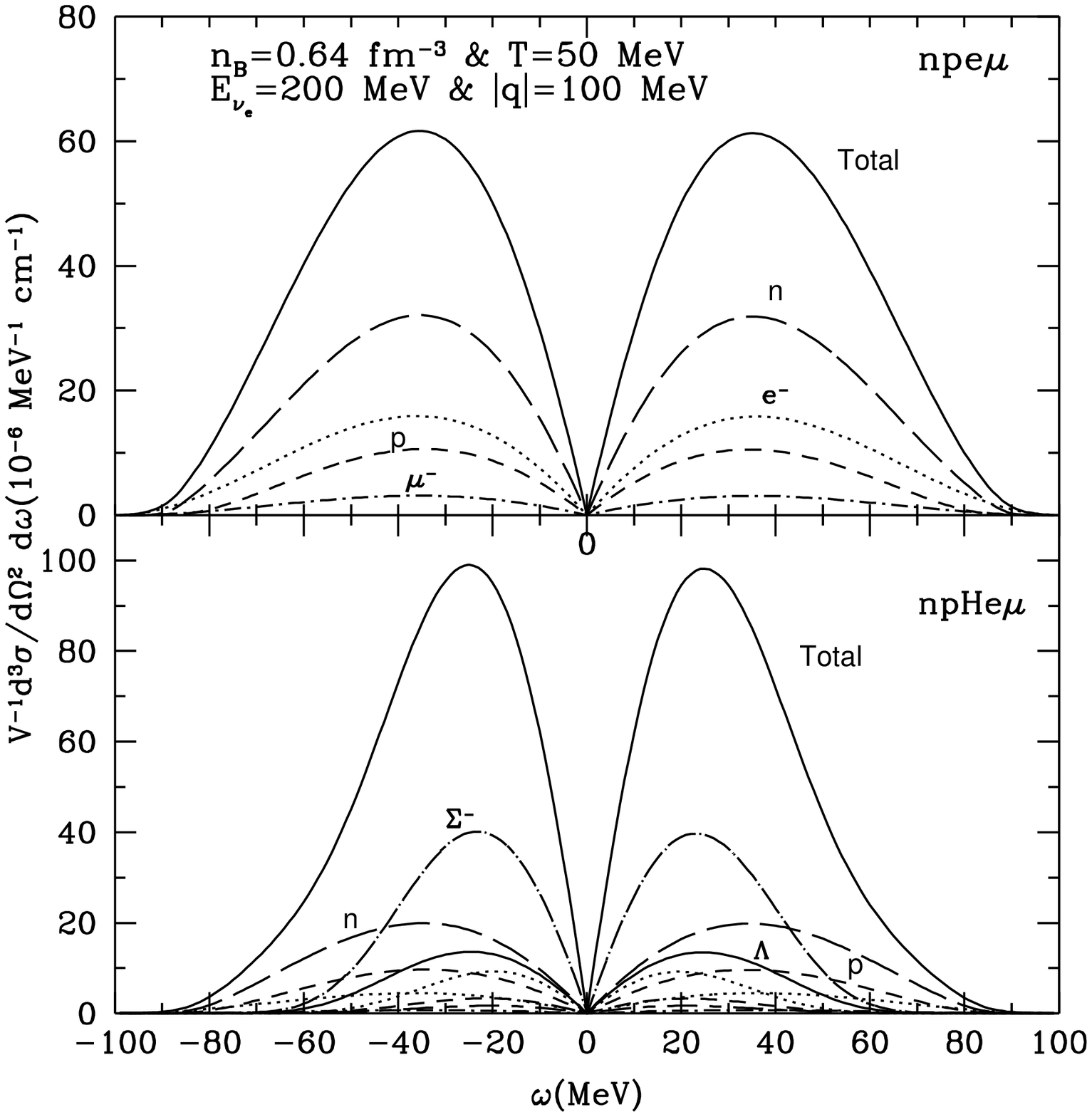}
\end{center}
\caption[]{\footnotesize }
{\label{cds}}
\end{figure}

In the cooling phase, neutrinos in the interior may be assumed to be  in
thermal equilibrium with Fermi-Dirac  distributions and zero chemical
potential; a typical thermal energy is $E_{\nu_e} \sim \pi T$. Thus, a
representative energy for thermal neutrinos  lies in the range 100--200 MeV. 
In Figure \ref{cds},   differential cross sections are shown at a baryon number
density of   $n_B=0.64~{\rm fm^{-3}}$ for $E_{\nu_e}=200$ MeV.  Scattering from
the $\Sigma^{-}$ hyperons dominates the contributions from the other particle
species in matter at small $\omega$.   
                             
\begin{figure}
\begin{center}
\leavevmode
\epsfxsize=6.0in
\epsfysize=6.0in
\epsffile{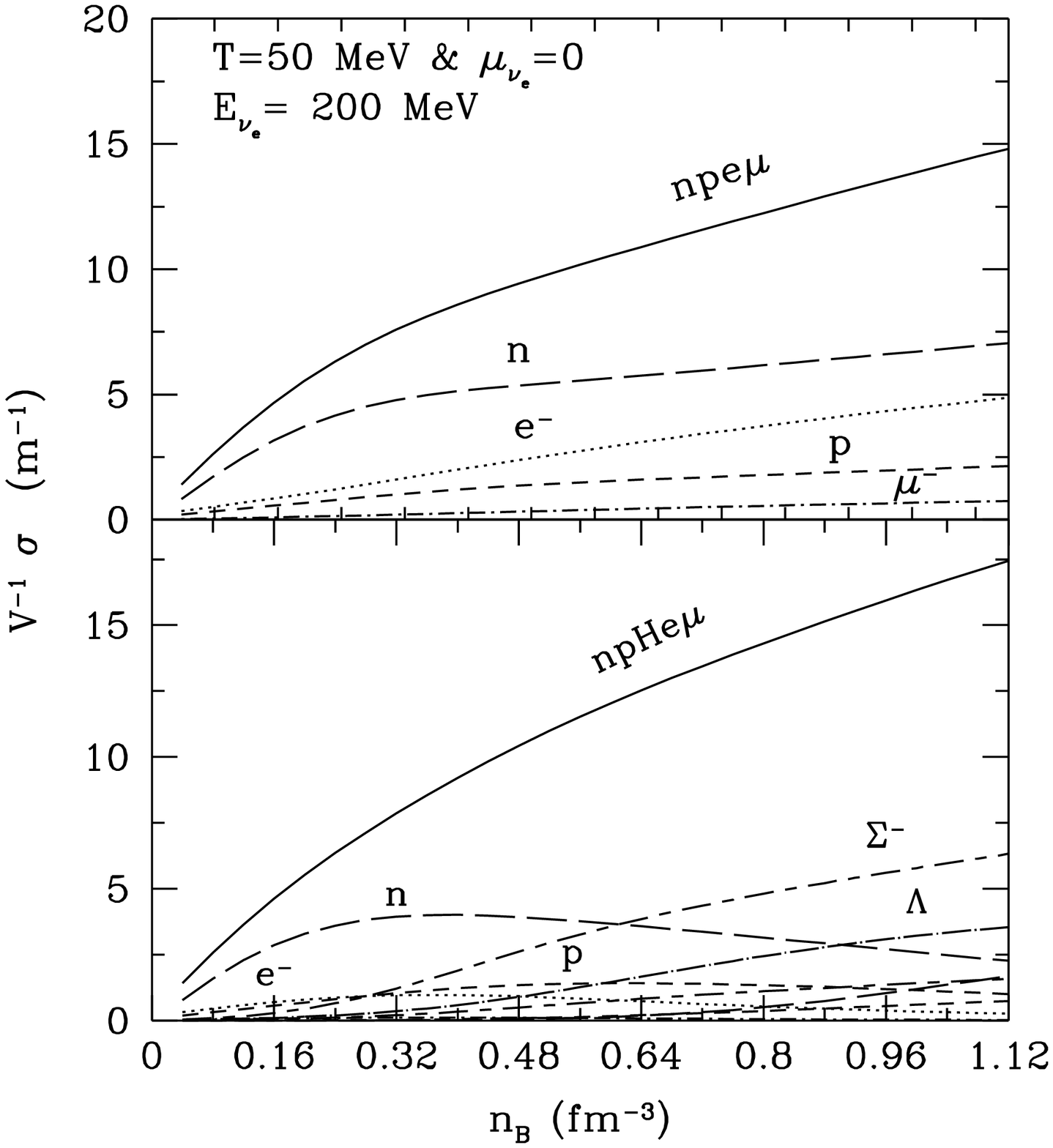}
\end{center}
\caption[]{\footnotesize }
{\label{sigcd}}
\end{figure}

The total cross-section per unit volume or the inverse collision mean free 
path is shown in Figure \ref{sigcd} for $E_{\nu_e}=200$ MeV  in matter with and
without hyperons. The results highlight the density dependence of the cross 
sections.   Here, since the electron neutrino chemical potential is zero,
neutrino final state blocking is unimportant.   The presence of hyperons in the
cooling phase  renders the interior more opaque to neutrinos relative to
nucleons only matter.   In fact, contributions from  hyperons dominate the
total cross section at high density.  Clearly, composition can  play an
important role in neutrino scattering also during the cooling phase. 
                                                                        
\begin{figure}
\begin{center}
\leavevmode
\epsfxsize=6.0in
\epsfysize=6.0in
\epsffile{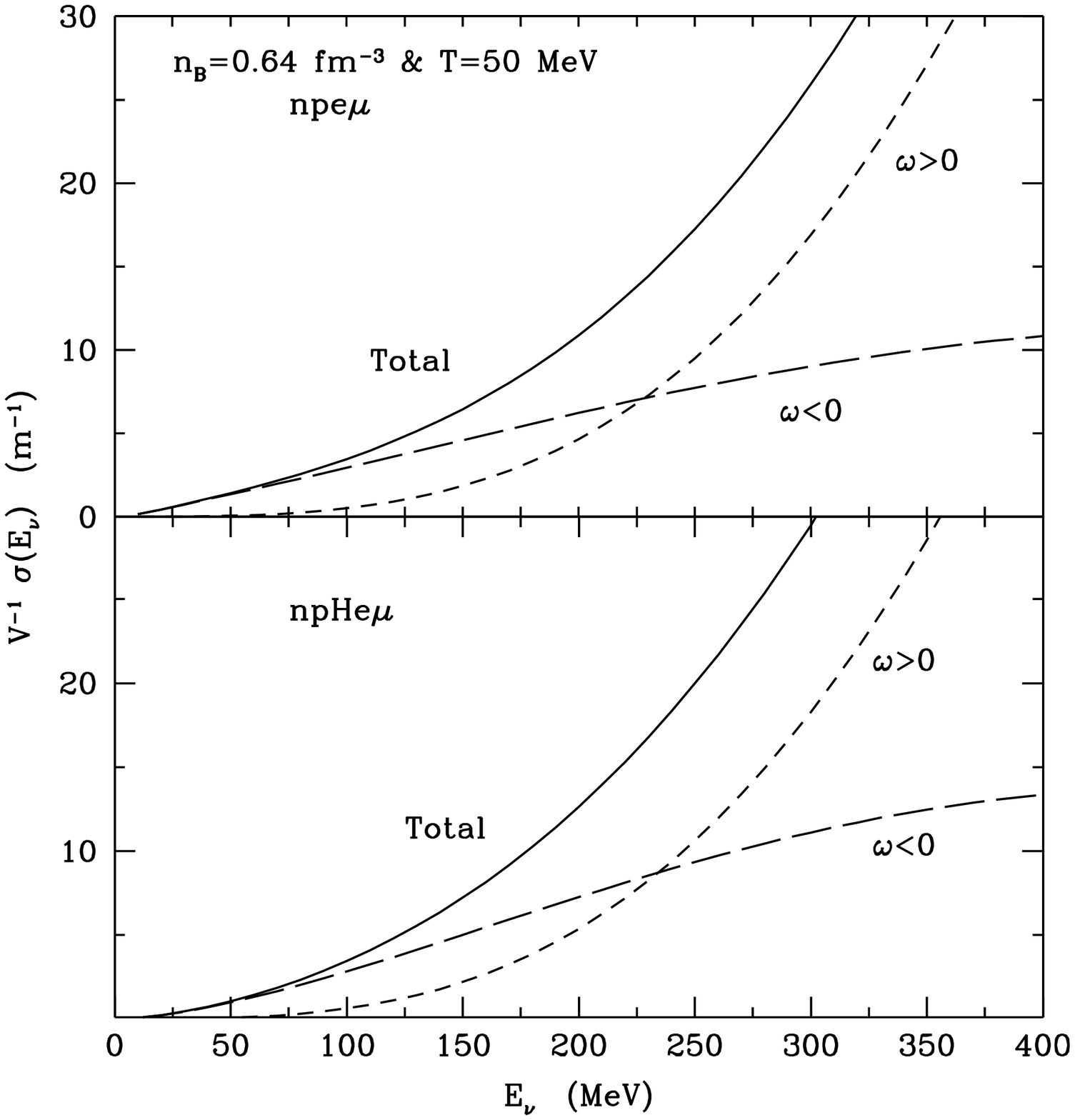}
\end{center}
\caption[]{\footnotesize }
{\label{sigce}}
\end{figure}

Unlike in the deleptonization phase,   where only neutrinos close to the Fermi
surface contribute to diffusion of lepton number, the transport of energy in the
cooling phase is determined by the cross sections of neutrinos of all energies.
To understand the energy  dependence, it is instructive to consider the
contributions from  negative $\omega$ (neutrino gains energy) and positive
$\omega$ (neutrino loses energy) separately (see Figure \ref{sigce}).   The
cross section grows linearly for scattering when the  neutrinos gain energy,
while the cross section grows approximately as $E^3$ for the case in which the
neutrinos lose energy.  It is important to note that, for thermal  neutrinos,
both contributions are important. The negative  $\omega$ phase space dominates
for low energy neutrino $(E_{\nu}\le \pi T)$ scattering, while the positive
$\omega$ phase space dominates at high energy $(E_{\nu}\ge \pi T)$.  The
precise form of the energy dependence of the neutrino scattering rates is
important in calculations of the energy averaged mean free paths for thermal
neutrinos (Prakash et al., 1996).  Our results suggest that an appropriate
parametrization would be                                
\begin{eqnarray}
V^{-1}\sigma(E)= A(n_B,T)E+B(n_B,T)E^3 
\end{eqnarray} 
for the energy dependence for thermal neutrinos in the interior regions. 

A detailed discussion of the various averaging schemes employed  in more
complete treatments of neutrino transport is beyond the scope of this work. We
hope to address these and related issues at a later time.

\section{CONCLUSIONS }

Our aim here has been to elucidate the effects of composition and of strong
interactions of the ambient matter on neutral current neutrino scattering 
cross sections during the deleptonization and cooling phases of the evolution
of a newly born neutron star.  Towards this end, we have calculated the
neutrino scattering  cross sections in protoneutron star matter, whose
constituents exhibit varying degrees of degeneracy during the evolution of the
star.   In both phases, the composition of matter is chiefly determined by the
nature of strong interactions and whether or not neutrinos are trapped in
matter.  In addition to the standard scenario, in which the strongly
interacting particles are only nucleons, we have explored the influence of  the
possible presence of  strangeness-bearing hyperons on the neutrino scattering
cross sections.  An important feature of our calculations is that the neutrino
opacities are consistent with the equation of state of matter at finite
temperature and density. 

We have identified neutral current neutrino interactions with hyperons that are
important sources of opacity.   Significant contribution to the neutrino
opacity arises from scattering involving  negatively charged  $\Sigma^-$ and
$\Xi^-$ hyperons, chiefly due to their large vector couplings, $|C_V| \sim 2$. 
Although the  contributions from the  neutral $\Lambda$ and $\Sigma^0$ are
smaller than those of nucleons, these particles, when present, furnish baryon
number which decreases the relative concentrations of nucleons.  This leads to
a larger opacity relative to nucleons only matter.  The neutrino cross sections
depend sensitively on the Fermi momenta and effective masses of the various
particles present in matter.  Whether or not a particular hyperon is present
depends on the many-body description of charge neutral beta-equilibrated
matter.  We find that as long as one or the other hyperon is present, the cross
sections are significantly modified from the case of nucleons only matter.   
                                                                    
In the deleptonization phase (lepton number fraction $Y_{Le}=0.4$ and $T \sim
20-30$ MeV), electron abundances are significantly larger than those in a cold
catalyzed star, since neutrinos are trapped in matter, whether or not hyperons
are present.  Consequently, electrons contribute nearly as much as neutrons to
the opacities.  In neutrino trapped matter, the appearance of negatively
charged hyperons (e.g., $\Sigma^-$) is delayed to higher densities (relative to
neutrino-free matter); also, their abundances are suppressed.  However, the
presence of neutral hyperons, such as the $\Lambda$, results in neutrino
abundances that grow with density. This leads to significant  enhancements in
the cross sections for neutrinos (of characteristic energies close to the local
neutrino chemical potential) compared to those in normal nucleonic matter. 
                    
Modifications due to strangeness in the cooling phase are quantitatively 
different from those in the deleptonization phase.   In the cooling phase,
in which matter is nearly neutrino free, the response of the $\Sigma^-$
hyperons to thermal neutrinos is the most significant.  Although a comparable
number of $\Lambda$ hyperons are present, neutrinos couple weakly to this
species, and, hence, their contributions are significant only at high density. 

Our findings here suggest several directions for further study.  The extension
to include correlations between the different particles and RPA corrections to
the results obtained here is under progress.  The presence of charged  
particles, such as the $\Sigma^-$, could make available low energy collective 
plasma modes through electromagnetic correlations, in addition to the scalar, 
vector and iso-vector correlations.  Calculations of neutrino opacities from 
charged current reactions (which are important during the deleptonization
phase), in strangeness-rich matter whose constituents exhibit varying degrees
of degeneracy, are required for a complete description of the evolution.  This
will be taken up in a separate work. Effects of strangeness on lepton number 
and energy transport may be studied by employing energy averages  (Rosseland
means) of the opacities  in present protoneutron codes.  Useful  tables of such
average opacities will be made generally available.  With new  generation
neutrino detectors capable of recording thousands of  neutrino  events, it may
be possible to distinguish between different scenarios  observationally.

\newpage
 
\appendix{ APPENDIX A}

The integral form of the imaginary parts of the various polarizations
that characterize the system's response to the neutrinos are collected here. 
The causal component of the density dependent polarizations in
Eqs.~(\ref{r1}-\ref{r3}) are given by Saito et al (1989).  For space like
excitations $(q_0\le |q|)$ and $q_{\mu}^2\le0$, they are given by 
\begin{eqnarray}
{\rm Im}~ \Pi_L(q_0,\vec{q})&=&\pi\lambda\int\frac{d^3p}{(2\pi)^3}~~
\frac{E_p^{*2}-|p|^2\cos^2\theta}{E_p^*E_{p+q}^*}~~\Theta^+ \,, \\
{\rm Im}~ \Pi_T(q_0,\vec{q})&=&\frac{\pi}{2}\lambda\int\frac{d^3p}{(2\pi)^3}~~
\frac{q_{\mu}^2/2-|p|^2(1-\cos^2\theta)}{E_p^*E_{p+q}^*}~~\Theta^+ \,,\\
{\rm Im}~ \Pi_S(q_0,\vec{q})&=&-\pi\lambda\int\frac{d^3p}{(2\pi)^3}~~
\frac{M^{*2}-q_{\mu}^2/4}{E_p^*E_{p+q}^*}~~\Theta^+ \,,\\
{\rm Im}~ \Pi_M(q_0,\vec{q})&=&\frac{\pi}{2}\lambda\int\frac{d^3p}{(2\pi)^3}~~
\frac{M^{*2}}{E_p^*E_{p+q}^*}~~\Theta^- \,, 
\end{eqnarray}
where
\begin{eqnarray}
\Theta^{\pm}&=&F^{\pm}(E_p^*,E_{p+q}^*)~[\delta(q_0-(E_{p+q}^*-E_p^*))+
\delta(q_0-(E_p^*-E_{p+q}^*))] \,,\\
F^{\pm}(E,E^*)&=&f_+(E)(1-f_+(E^*) \pm f_-(E)(1-f_-(E^*) \,,\\
E_{p}^*&=&\sqrt{|p|^2+M^{*2}} \,.\\
\end{eqnarray}
Above, $\lambda=2$ is the spin degeneracy factor. The particle distribution
functions $f_{\pm}(E^*)$ are the  Fermi-Dirac distribution functions     
\begin{equation}
f_{\pm}(E^*)=\frac{1}{1+\exp(\frac{E^*\mp\nu}{kT})}\,,
\end{equation}
where $\nu $ is the effective chemical potential defined in Eq.~(\ref{hyp3}). 
The angular integrals are performed by exploiting the delta functions, and the 
three dimensional integrals can be reduced to the following one dimensional 
integrals: 
\begin{eqnarray}
{\rm Im}~ \Pi_L(q_0,\vec{q})&=&\frac{\lambda}{4\pi}\frac{q_{\mu}^2}{|q|^3}
\int_{e_-}^{\infty}dE~~[(E+q_0/2)^2-|q|^2/4]\nonumber\\ 
&\times&[F^+(E,E+q_0)+F^+(E+q_0,E)] \,,\\
{\rm Im}~ \Pi_T(q_0,\vec{q})&=& \frac{\lambda}{8\pi}
\frac{q_{\mu}^2}{|q|^3}\int_{e_-}^{\infty}dE~~
[(E^*+q_0)^2+|q|^2/4+|q|^2M^{*2}/q_{\mu}^2)]\nonumber\\
&\times & [F^+(E,E+q_0)+F^+(E+q_0,E)] \,,\\
{\rm Im}~ \Pi_S(q_0,\vec{q})&=&-\frac{\lambda}{4\pi|q|}(M^{*2}-q_{\mu}^2/4)
\int_{e_-}^{\infty}dE~~[F^+(E,E+q_0)+F^+(E+q_0,E)] \,,\\
{\rm Im}~ \Pi_M(q_0,\vec{q})&=&\frac{\lambda M^*}{8\pi |q|}
\int_{e_-}^{\infty}dE~~[2E+q_0][F^-(E,E+q_0)+F^-(E+q_0,E)] \,.
\end{eqnarray}
The axial and the vector-axial polarizations entering the neutrino
differential cross-sections are related to the scalar and mixed polarizations 
defined above through the following relations
\begin{eqnarray}
{\rm Im}~ \Pi_A(q_0,\vec{q})&=&\frac{M^{*2}}{q_{\mu}^2/4-M^{*2}}{\rm Im}~ \Pi_S
(q_0,\vec{q}) \,,\\
{\rm Im}~ \Pi_{VA}(q_0,\vec{q})&=&\frac{q_{\mu}^2}{2|q|^2M^*}{\rm Im}~ \Pi_{M}
(q_0,\vec{q}) \,.
\end{eqnarray}
The causal polarizations are related to the retarded or time ordered 
polarizations through
\begin{equation}
{\rm Im}~\Pi^R(q_0,q)=\tanh\left(\frac{-q_0}{2kT}\right){\rm Im}~\Pi^C(q_0,q)
\,.
\end{equation}
The phase space integrals 
\begin{equation}
I_n=\tanh\left(\frac{-q_0}{2kT}\right)
\int_{e_-}^{\infty}dE~E^n~[F(E,E+q_0)+F(E+q_0,E)] 
\label{pspace}
\end{equation}
simplify the representation of the retarded polarizations.  
The one dimensional phase space integrals are explicitly evaluated and 
expressed in terms of polylogaritmic functions.  This representation
is particularly useful in understanding the behavior with density and
temperature and is given in Eq. (\ref{ins}) in \S 2. 
For our purpose, only the three integrals $I_0,~I_1$, and $I_3$ are required.
Their analytical representations are also given in \S 2.
\vspace*{0.2in}

\vspace*{0.3in} 

This work was supported in part by the U.S. Department of Energy under
contract number DOE/DE-FG02-88ER-40388 and by the NASA grant NAG 52863.  We
thank Jim Lattimer for helpful discussions and for a careful reading of the
paper.  We are grateful to David Kaplan and Martin Savage for their help 
in the preparation of Table I. 

\newpage

\centerline {REFERENCES} 
\begin{tabbing}
xxxxx\=xxxxxxxxxxxxxxxxxxxxxxxxxxxxxxxxxxxxxxxxxxxxxxxxxxxxxxxxxxxxxxxxxxxxxxxx\kill
\ni Bionta, R. M., et al., 1987, Phys. Rev. Lett. 58, 1494 \\
\ni Bruenn, S., 1985, ApJ. Suppl. 58, 771 \\
\ni Burrows, A., 1988, ApJ 334, 891 \\
\ni -------, 1990, Ann. Rev. Nucl. Sci. 40, 181 \\ 
\ni Burrows, A \& Lattimer, J. M., 1986, ApJ 307, 178 \\
\ni Burrows, A \& Mazurek, T. J., 1982, ApJ 259, 330 \\
\ni Cooperstein, J., 1988, Phys. Rep. 163, 95 \\
\ni Gaillard, J.-M., \& Sauvage, G. 1984, Ann. Rev. Nucl. Part. Sci., 34, 351
\\
\ni Ellis, J., Kapusta, J. I., \& Olive, K. A. 1991, 
Nucl. Phys. B348, 345 \\
\ni Ellis, P. J., Knorren, R., \& Prakash, M., 1995, Phys. Lett. B349, 11 \\ 
\ni Fetter, A. L., \& Walecka, J. D., 1971, Quantum Theory of Many-Particle
Systems \\
\ni -------(New York: McGraw-Hill) \\
\ni Glashow, S. L., 1961, Nucl. Phys. 22, 579 \\
\ni Glendenning, N. K., \& Moszkowski, S. A. 1991, 
Phys. Rev. Lett. 67, 2414 \\
\ni Goodwin, B. T., 1982, ApJ 261, 321 \\ 
\ni Goodwin, B. T. \& Pethick, C. J. 1982, ApJ 253, 816 \\ 
\ni Hirata, K., et al., 1987, Phys. Rev. Lett. 58, 1490 \\
\ni Horowitz, C. J., 1992, Phys. Rev. Lett. 69, 2627 \\
\ni Horowitz, C. J., \& Serot, H. S. 1981, Nucl. Phys. A368, 503 \\
\ni Horowitz, C. J., \& Wehrberger, K., 1991a, Nucl. Phys. A531, 665 \\
\ni --------, 1991b, Phys. Rev. Lett. 66, 272 \\
\ni --------, 1992, Phys. Lett. B226, 236 \\
\ni Iwamoto, N. 1982, Ann. Phys. 141, 1                           \\
\ni Iwamoto, N., \& Pethick, C. J., 1982, Phys. Rev. D25, 313 \\
\ni Kaplan, D. B., \& Nelson, A. E. 1986, Phys. Lett. B175, 57      \\
\ni ------, 1986, Phys. Lett. B179, 409 (E)                            \\
\ni Kapusta, J. A., \& Olive, K. A., 1990, Phys. Rev. Lett. 64, 13      \\
\ni Keil, W., \& Janka, H. T., 1995, Astron. \& Astrophys. 296, 145 \\
\ni Knorren, R., Prakash, M., \& Ellis, P. J., 1995, Phys. Rev. C52, 3470 \\
\ni Lamb, D. Q., 1978, Phys. Rev. Lett. 41, 1623 \\ 
\ni Lamb, D. Q., \& Pethick, C. J., 1976, ApJ 209, L77 \\ 
\ni Lewin, L., 1983, Polylogarithms and Associated Functions 
(New York: North-Holland) \\
\ni Lim, K., \& Horowitz, C. J., 1989, Nucl. Phys. A501, 729 \\ 
\ni Mazurek, T. J., 1975, Astrophys. Space. Sci. 35, 117 \\
\ni Mare\v{s}, J., Friedman, E., Gal. A., \& Jennings, B. K., 1995, 
Nucl. Phys. A594, 311 \\ 
\ni Maxwell, O., 1987, ApJ 316, 691 \\
\ni Prakash, M., Prakash, Manju, Lattimer, J. M., \& Pethick, C. J.
1992, ApJ 390, L77 \\
\ni Prakash, M., Cooke, J., \& Lattimer, J. M., 1995, Phys. Rev. D52, 661 \\
\ni Prakash, M., Bombaci, I., Prakash, Manju, Ellis, P. J., \\
\ni -------Lattimer, J. M., \& Knorren, R., 1996, Phys. Rep., In press \\
\ni Reddy, S., and Prakash, M., 1995, Proc. of the 11th Winter Workshop on
Nuclear \\ \> Dynamics, Key West, Fl, Feb 11--18, 1995 \\
\ni Sato, K., 1975, Prog. Theor. Phys. 53, 595 \\
\ni Saito, K., Marayuma, T., \& Soutame, K., 1989, Phys. Rev. C40, 407 \\
\ni Salam, A., 1968, Elementary Particle Theory: Relativistic Groups and
Analyticity, \\ 
\ni ------ Proceedings of the Eighth Nobel Symposium, \\ 
\ni ------ ed. N. Svartholm (Stockholm: Almquist \& Wiskell) \\
\ni Sawyer, R. F., 1975, Phys. Rev. D11, 2740 \\
\ni ------, 1989, Phys. Rev. C40, 865 \\
\ni ------, 1995, Phys. Rev. Lett. C75, 2260 \\
\ni Sawyer, R. F. \& Soni, R., 1979, ApJ 230, 859 \\
\ni Serot, B. D., \& Walecka, J. D., 1986, in Adv. Nucl. Phys. 16, \\ 
ed. J. W. Negele \& E. Vogt (New York: Plenum) \\ 
\ni Thorsson, V., Prakash, M., \& Lattimer, J. M. 1994, Nucl. Phys. A572, 693
\\
\ni Tubbs, D. L., \& Schramm, D. N., 1975, ApJ 201, 467 \\
\ni van den Horn. L. J., \& Cooperstein, J., 1986, ApJ 300, 142 \\
\ni Weinberg. S., 1967, Phys. Rev. Lett. 19, 1264 \\
\end{tabbing}

\newpage
\centerline {FIGURE CAPTIONS}
\vspace*{0.2in} 

\ni Fig. 1--Particle fractions, $Y_i=n_i/n_b$ at zero temperature, 
for the model GM1. 
The upper panel refers to nucleons-only matter. The lower panel shows results
in matter with hyperons. \\                       

\ni Fig. 2--Particle fractions in the deleptonization phase ($T=20$ MeV). The
lepton fraction  $Y_{Le}=Y_{\nu_e}+Y_{e}$ is chosen to be 0.4. \\ 

\ni Fig. 3--Particle fractions in the cooling phase 
($T=50$ MeV and $Y_{\nu_e}=0$). \\                       

\ni Fig. 4--Neutrino differential scattering cross sections in pure  neutron
matter at $T=10$ MeV. Curves labelled 1 and 2 refer to baryon densities of 
$n_B=0.32 {\rm ~fm^{-3}}$ and $n_B=0.64 {\rm ~fm^{-3}}$, respectively.  Curves
labelled 1(a) and 2(a) are results with interactions as in a mean field
theoretical model. Those labelled 1(b) and 2(b) are results for a free Fermi
gas. \\                       
                                 
\ni Fig. 5--Neutrino differential scattering cross sections in pure  neutron
matter at $T=10,20$, and 50 MeV. Interactions as in a mean
field theoretical model are included.  Upper panel results are for 
$n_B=0.16 {\rm ~fm^{-3}}$ and lower panel results are for $n_B=0.64{\rm
~fm^{-3}}$. \\                         
                                                                  
\ni Fig. 6--Neutrino differential scattering cross sections in charge
equilibrated matter at zero temperature. Upper panel results are for matter
with nucleons only.  The lower panel shows results in matter with hyperons.\\
           
\ni Fig. 7--Same as in Figure 6, but for charge equilibrated matter at 
$T=30$ MeV. \\                       
                        
\ni Fig. 8--Neutrino differential scattering cross sections in charge
equilibrated neutrino-trapped matter in the deleptonization phase. 
Upper panel results are for matter
with nucleons only.  The lower panel shows results in matter with hyperons.\\

\ni Fig. 9-- Neutrino total scattering cross sections versus neutrino energy 
in charge equilibrated neutrino-trapped matter in the deleptonization phase. 
Blocking of final state neutrinos is included. \\

\ni Fig. 10--Upper panel: Neutrino total scattering cross sections in charge
equilibrated neutrino-trapped matter in the deleptonization phase.  The
neutrino energy is set equal to the local neutrino chemical potential at each
density.  Lower panel: Electron
neutrino chemical potential. \\ 

\ni Fig. 11--Neutrino differential scattering cross sections in charge
equilibrated neutrino-free matter in the cooling phase. 
Upper panel results are for matter
with nucleons only.  The lower panel shows results in matter with hyperons.\\ 

\ni Fig. 12--Neutrino total scattering cross sections in charge
equilibrated neutrino-free matter in the cooling phase. Important individual
contributions to the total cross sections are also shown. \\                  
    
\newpage
       
\ni Fig. 13--Energy dependence of the neutrino total scattering 
cross sections in charge equilibrated neutrino-free matter in the 
cooling phase.  Contributions from the positive energy transfer and the 
negative energy transfer processes, and their sum are shown in matter with
nucleons only (upper panel) and in matter with hyperons (lower panel). \\

\newpage
\centerline {TABLE CAPTIONS}

\ni TABLE 1. NOTE.-- Coupling constants derived assuming SU(3) symmetry and the
constituent quark model for the hadrons.  Numerical values are quoted using 
D=0.756 , F=0.477, $\sin^2\theta_W$=0.23 and $\sin\theta_c = 0.231$ (Gaillard
\& Sauvage 1984). \\

\ni TABLE 2. NOTE.-- Constants from Glendenning and Moszkowski (1991). \\

\newpage
\vspace*{0.2in}

\begin{center}
\vskip 10pt
\centerline{TABLE 1}
\vskip 5pt
\centerline{NEUTRAL CURRENT VECTOR AND AXIAL COUPLINGS}
\vskip 5pt
\begin{tabular}{lcc}
\hline \hline
{Reaction } & $C_V$  & $C_A $  \\
\hline 
$ \nu_e + e \rightarrow \nu_e + e$ & $ 1+4\sin^2\theta_W=1.92$ & $1$\\
$\nu_e + \mu \rightarrow \nu_e + \mu$ & $ -1+4\sin^2\theta_W=-0.08$ & $-1$\\
$\nu_i + n \rightarrow \nu_i + n$ & $-1$ & $-D-F=-1.23$ \\
$\nu_i + p \rightarrow \nu_i+ p$ & $ 1-4\sin^2\theta_W=0.08$ & $D+F=1.23$
  \\
$\nu_i + \Lambda\rightarrow \nu_i +\Lambda$ & $-1$ & $-F-D/3=-0.73 $ \\
$\nu_i + \Sigma^-\rightarrow \nu_i +\Sigma^-$ &$ -3+4\sin^2\theta_W=-2.08$ &
 $D-3F=-0.68$  \\
$\nu_i + \Sigma^+\rightarrow \nu_i +\Sigma^+$&$ 1-4\sin^2\theta_W=0.08$ & D+F
=1.23 \\
$\nu_i + \Sigma^0\rightarrow \nu_i +\Sigma^0$ & $-1$ & $D-F=0.28$\\
$\nu_i + \Xi^-\rightarrow \nu_i +\Xi^-$ & $ -3+4\sin^2\theta_W=-2.08 $ & $D-3F
=-0.68$ \\
$\nu_i + \Xi^0\rightarrow \nu_i +\Xi^0$ & $-1$ & $-D-F=-1.23$ \\
$\nu_i+\Sigma^0 \rightarrow \nu_i+\Lambda$ & $0$ & $2D/\sqrt 3=0.87$
 \\ \hline 
\end{tabular}
\vskip 5pt

\end{center}

\newpage

\vskip 10pt
\centerline {TABLE 2}
\vskip 5pt
\centerline{NUCLEON-MESON COUPLING CONSTANTS  }
\begin{center}
\begin{tabular}{ccccccc}
\hline \hline 
Model & $\frac{g_{\sigma}}{m_{\sigma}}$ 
&$\frac{g_{\omega}}{m_{\omega}}$&$\frac{g_{\rho}}{m_{\rho}}$& b & c& 
$\frac{M^*}{M}$ \\ 
 &(fm)  &(fm)  &(fm)  &  &   & \\
\hline 
GM1 & 3.434 &2.674  &2.100  &0.00295  &$-$0.00107   &0.70 \\
\hline  
\end{tabular}
\end{center}

\newpage

\vskip 10pt

\vskip 10pt

\vskip 10pt

\end{document}